\documentclass[manuscript,a4paper,article]{aastex}

\newcommand{\Chandra}{{\it Chandra}}

\newcommand{\XMM}{{\it XMM-Newton}}

\newcommand{\ROSAT}{{\it ROSAT}}
\newcommand{\Einstein}{{\it Einstein}}
\newcommand{\ASCA}{{\it ASCA}}

\usepackage{epsfig}
\usepackage{amsmath}
\usepackage[breaklinks]{hyperref}
\usepackage{multirow}
\usepackage{color}

\shorttitle{}
\shortauthors{Hu et al.}

\begin{document}

\title{The Merger Dynamics of the Galaxy Cluster Abell~1775: New Insights from \Chandra\ and \XMM\ for a Cluster Simultaneously Hosting a WAT and a NAT Radio Sources}

\author{\sc Dan Hu\altaffilmark{1}, Haiguang Xu\altaffilmark{1,2}, Zhenghao Zhu\altaffilmark{1}, Chenxi Shan\altaffilmark{1}, Yongkai Zhu\altaffilmark{1}, Shida Fan\altaffilmark{1}, Yuanyuan Zhao\altaffilmark{1}, Chengze Liu\altaffilmark{1}, Hoongwah Siew\altaffilmark{1}, Zhongli Zhang\altaffilmark{3}, Liyi Gu\altaffilmark{4}, Melanie Johnston-Hollitt\altaffilmark{5}, Xi Kang\altaffilmark{6}, Qinghua Tan\altaffilmark{7}, Jiang Chang\altaffilmark{7}, and Xiang-ping Wu\altaffilmark{8}
}

\altaffiltext{1}{School of Physics and Astronomy, Shanghai Jiao Tong University, 800 Dongchuan Road, Minhang, Shanghai 200240, China; email: hudan\_bazhaoyu@sjtu.edu.cn, hgxu@sjtu.edu.cn}
\altaffiltext{2}{IFSA Collaborative Innovation Center, Shanghai Jiao Tong University, Minhang, Shanghai 200240, China}
\altaffiltext{3}{Shanghai Astronomical Observatory, Chinese Academy of Sciences, 80 Nandan Road, Shanghai 200030, China}
\altaffiltext{4}{SRON Netherlands Institute for Space Research, Sorbonnelaan 2, 3584 CA Utrecht, the Netherlands}
\altaffiltext{5}{International Centre for Radio Astronomy Research (ICRAR), Curtin University, Bentley, WA 6102, Australia}
\altaffiltext{6}{Zhejiang University-Purple Mountain Observatory Joint Research Center for Astronomy, Zhejiang University, Hangzhou 310027, China}
\altaffiltext{6}{Purple Mountain Observatory, Chinese Academy of Sciences, No.10 Yuanhua Road, Qixia District, Nanjing 210034, China}
\altaffiltext{8}{National Astronomical Observatories, Chinese Academy of Sciences, 20A Datun Road, Beijing 100012, China}

\begin{abstract}

We present a new study of the merger dynamics of Abell~1775 by analyzing the high-quality \Chandra\ and \XMM\ archival data. We \textbf{confirm/identify} an arc-shaped edge (i.e., the head) at $\sim48$~kpc west of the X-ray peak, a split cold gas tail that extends eastward to $\sim163$~kpc, and a plume of spiral-like X-ray excess (within about $81-324$~kpc northeast of the cluster core) that connects to the end of the tail. The head, across which the projected gas temperature rises outward from $3.39_{-0.18}^{+0.28}$~keV to $5.30_{-0.43}^{+0.54}$~keV, is found to be a cold front with a Mach number of $\mathcal{M}\sim0.79$. Along the surfaces of the cold front and tail, typical KHI features (noses and wings, etc.) are found and are used to constrain the upper limit of the magnetic field ($\sim11.2~\mu$G) and the viscosity suppression factor ($\sim0.01$). \textbf{Combining optical and radio evidence we propose a two-body merger (instead of systematic motion in a large-scale gas environment) scenario and have carried out idealized hydrodynamic simulations to verify it.} We find that the observed X-ray emission and temperature distributions can be best reproduced with a merger mass ratio of 5 after the first pericentric passage. The NAT radio galaxy is \textbf{thus} more likely to be a single galaxy falling into the cluster center at a relative velocity of 2800~$\rm km~s^{-1}$, \textbf{a speed} constrained by its radio morphology. \textbf{The infalling subcluster is expected to have a relatively low gas content, because only a gas-poor subcluster can cause central-only disturbances as observed in such an off-axis merger.} 
%The observed substructures can only be reproduced when the subcluster has a relatively low gas content, which is likely to be the reason for the phenomena of central-only disturbances during the off-axis merger.
\end{abstract}


%#####################
\section{INTRODUCTION}
%#####################

Within the frame of standard dark matter-driven hierarchical structure formation theory, galaxy clusters and groups, the largest celestial objects in gravitationally bound state in our universe, are believed to form via a sequence of mergers (when the mass change is $\Delta m > 10^{12}~M_{\sun}$; e.g., \citealt{CB05}) and accretion events ($\Delta m \le 10^{12}~M_{\sun}$) of subsystems. During typical merging processes a variety of characteristic features such as cold fronts, shocks, radio halos, radio relics, as well as head-tail radio sources (see, e.g., \citealt{begelman84}, \citealt{MV07} and \citealt{feretti12} for reviews) can form in different stages under proper conditions, which largely depend on the initial conditions of the merger (the mass ratio of the merging clusters, initial relative velocity, impact parameter, etc.; e.g., \citealt{KB12}).
These features, which are often seen in the high-resolution X-ray and radio images of the merging systems, provide us with invaluable information about the physics and history of the merger. Therefore, they can be utilized as a tight constraint in theoretical and numerical studies of the evolution history of galaxy clusters and groups \citep[e.g.,][]{springel07,ML13,hu19}.

Cold fronts, i.e., sharp subsonic discontinuities in the intra-cluster medium (ICM), across which gas density drops abruptly while gas temperature rises, are commonly found in merging galaxy clusters, galaxy groups, and even elliptical galaxies. As one of the most important \Chandra\ discoveries (see \citealt{markevitch00} and \citealt{vikhlinin01} for the first detections in Abell~2142 and Abell~3667, respectively), they provide a vital probe for studying the merger dynamics (see \citealt{MV07} and \citealt{ZR16}, for reviews). Cold fronts are usually classified into three categories.
First, when a subcluster (sometimes a single galaxy) falls into a cluster at a relatively high speed, a leading edge can form as the boundary of the subcluster's cool core to separate the cool core from the hot environment. Because this type of leading edges are often accompanied by a gas tail stripped from the subcluster's core due to the ram pressure force, they are classified as the ``remnant core'' type cold fronts (or merger cold fronts). Typical examples are 1E0657$-$56 (the Bullet cluster; \citealt{markevitch02}), Abell~3667 \citep{vikhlinin01,datta14}, the RX~J0751.3+5012 group \citep{russell14}, and the infalling NGC~1404 galaxy in the Fornax cluster \citep{machacek05,su17}. 

Second, cold fronts were also serendipitously found in Abell~2142 \citep{markevitch00}, Abell~1795 \citep{markevitch01}, and RX~J1720.1+2638 \citep{mazzotta01} relaxed clusters, and later in dozens of clusters that show few or no apparent merger signatures. This type of cold fronts tends to reside in the core regions ($\lesssim 100 $~kpc) of cool core clusters and always exhibit a spiral pattern with spiral gas tails, which is speculated to have been caused by the gas sloshing process triggered by the angular momentum input during mergers with nonzero impact parameters \citep{TH05,AM06,poole06,SF07,zuhone11,hu19}; they are called ``sloshing'' type cold fronts. Systems hosting a sloshing type cold front include Abell~1795 \citep{markevitch01,ehlert15}, Abell~2204 \citep{sanders05}, the Ophiuchus cluster \citep{million10}, the HCG~62 group \citep{rafferty13,hu19}, etc. Actually, rough estimations based on sample analysis indicated that the sloshing type cold fronts may appear in the cool cores of $\sim 1/2-2/3$ relaxed galaxy clusters \citep[e.g.,][]{ghizzardi10}. 
Moreover, recent observations revealed that similar large-scale cold fronts also appear in cluster's skirt regions, usually at about $0.5-1$ virial radius (e.g., Abell~1763, \citealt{douglass18}; Abell~2142, \citealt{rossetti13}; RX~J2014.8-2430, \citealt{walker14}; the Perseus cluster, \citealt{simionescu12,walker18,walker20}), demonstrating that these giant cold fronts originate and survive in the cluster core and then propagate to outer regions on timescales of several Gyrs. 

\textbf{
However, the enormously cold fronts near the cluster virial radius have an alternative interpretation. By using spherical one-dimensional hydrodynamic simulations, \citet{BKH10} first proposed the possibility of a third type of cold fronts, namely “shock-induced” type cold front. When two shocks propagating along the same direction, the secondary shock might chase and merge with the primary shock (in most cases, the virial shock), resulting in a quasi-spherical contact discontinuity in both density and temperature. 
Recently, \citet{zhang20} carried out one-dimensional hydrodynamic simulations to reproduce the newly discovered cold front at the virial radius ($\sim 1.7$~Mpc) of the Perseus cluster \citep{walker20}. They found that this Mpc-scale cold front can be explained using the merger between the runaway merger shock and the virial shock.}

In addition to the cold fronts and gas tails, another characteristic feature is that their boundaries can be distorted by the Kelvin-Helmholtz instabilities (KHIs) caused by the shear flows,
a phenomenon that has been predicted in high-resolution hydrodynamic simulations in which the effects of both viscosity and magnetic field have not yet been considered \citep[e.g.,][]{heinz03,zuhone10,roediger11,roediger12b}, and has been observed in some clusters and groups in high-resolution X-ray observations. 
Corresponding substructures include the noses and wings in the RX~J0751.3$+$5012 group \citep{russell14} and the group pair NGC~7618$-$UGC~12491 \citep{roediger12b}, the KH eddies in Abell~3667 \citep{ichinohe17}, Abell~2142 \citep{WM18}, and the elliptical galaxy NGC~1404 \citep{su17}, the concave bays usually bracketed by the KH rolls in Perseus, Centaurus, and Abell~1795 \citep{walker17}, and the horns in the Virgo galaxy M89 \citep{roediger15}. However, in other clusters and groups with good image quality, the observed cold fronts and gas tails show relatively smooth morphologies. The absence of the KHIs in these sources is possibly caused by the sufficiently strong magnetic field with an orientation parallel to the shearing surface, which tends to stabilize the front \citep{VM02}, and/or by the local viscosity \citep{roediger13}. Numerical works \citep[e.g.,][]{DP08,zuhone10,ZML11,zuhone15,roediger12a,roediger13,roediger15,suzuki13} that incorporate the effects of either magnetic field, or viscosity, or both, have been performed to demonstrate how and to what extend the growth of KHIs is suppressed. The authors found that the simulated cold front morphologies apparently depend on the choice of magnetic field and viscosity on scales of tens of kpc \citep[see also][]{ZR16}. Therefore, the observed KHI substructures can be used as a powerful probe to constrain ICM physics.

In this work, we analyze the high-quality \Chandra\ and \XMM\ archive data to carry out a detailed X-ray imaging spectroscopic study of the galaxy cluster Abell~1775 ($z = 0.0717$), which is rich (richness class $R=2$), non-relaxed, and X-ray luminous ($L_{\rm X,~0.1-2.4~keV} = 2.91 \times 10^{44}$~$\rm ergs~s^{-1}$; \citealt{ebeling98}). 
The cluster was firstly found to contain two interacting subclusters by studying the distribution of the radial velocities of its 51 member galaxies \citep{OHF95}.
\citet{zhang11} further constructed an enlarged sample (151 member galaxies) by combining the photometric data derived from both the Sloan Digital Sky Survey (SDSS) and the Beijing-Arizona-Taiwan-Connecticut (BATC) multicolor photometry, and reported a spatial separation of $\sim 14\arcmin$ ($\sim 1.1$~Mpc) and a radial velocity difference of $\sim 2980$~$\rm km~s^{-1}$ between the two subclusters. 
In the X-ray band two X-ray gas halos were detected with both \Einstein\ and \ROSAT. The fainter one is found located about $\sim 1$~Mpc southeast to the main X-ray halo \citep{MKU89}, and both the X-ray halos spatially coinciding with the optical subclusters.
Meanwhile, the cluster is a rare case that simultaneously contains two tailed radio galaxies in its central region ($< 40$~kpc), which have been studied in detail using the Very Large Array (VLA), the Giant Metrewave Radio Telescope (GMRT), the Karl G. Jansky Very Large Array (JVLA), \textbf{and LOw Frequency ARray (LOFAR) in 144~MHz $-$15~GHz \citep{OO85,GF00,giacintucci07,terni17,botteon21}.} One of them shows relatively small ($\sim 10\arcsec$ or $\sim 14$~kpc) bent tails at 1.4~GHz \citep{OO85}, which point approximately toward the east. It is identified as a wide-angle tail (WAT) radio galaxy and is found to be associated with the brightest cluster galaxy (BCG) UGC~08669~NED01 (or SDSS~J134149.14+2622224.5; the peak of the optical main cluster), which appears as the central dominating elliptical galaxy. Located at about 28~kpc southeast of the BCG, \textbf{the other radio source exhibits an extremely long ($\sim$ 800~kpc) narrow tail at 144~MHz, which extends toward the northeast \citep{botteon21},} and is identified as a narrow-angle tail (NAT) radio galaxy. It is hosted by the giant elliptical galaxy UGC~08669~NED02 (or SDSS~J134150.45+262213.0), which possesses a similar luminosity to UGC~08669~NED01. \textbf{Besides, a small-scale radio halo in the central region of Abell~1775 was found recently in the 144~MHz LOFAR image, which indicates a merger event \citep{botteon21}.  } 

%[DELETE ! The appearance of WAT radio galaxies is usually regarded as a signature of an ongoing merger event \citep{burns94}. ]

However, the merger scenario of Abell~1775 is still unclear. Although the WAT hosted by the BCG belongs to the optical main cluster and its location accords with the X-ray peak of the main halo, the origin of the NAT radio galaxy is under debate (which optical subcluster does the NAT radio galaxy belong to, or rather, dose it belong to the third unrecognized substructure?).
No evidence for KHIs was reported in previous works. Nevertheless, based on the recent observations of Abell~3667 \citep{ichinohe17} and numerical simulations of viscous gas stripping of galaxies \citep{roediger15}, both of them showing a sharp cold front, a stripped tail, and signatures of the developing KHIs along the interface, we speculate that the KHI substructures are likely to occur and to be detected in Abell~1775 with the high-quality \Chandra\ data. Thus Abell~1775 appears to be a very interesting target for us to study the cluster merger process.

We list the basic properties of Abell~1775, as well as those of the two tailed radio galaxies in Table~\ref{tbl-1}. We organize this paper as follows: in Section~2 we describe the procedure of the \Chandra\ and \XMM\ data reduction; in Section~3 we extract and study the X-ray images and spectra; in Section~4 we utilize the motion of the gas core and KHI features to constrain the magnetic field and viscosity, discuss and discriminate between the possible merger scenarios based on our hydrodynamic simulations, and use the radio morphologies of two tailed radio galaxies to constrain their velocities relative to the ICM; in Section~5 we summarize our main conclusions.
Throughout this paper we quote errors at the $90\%$ confidence level unless otherwise stated. We adopt cosmological parameters $H_0 = 70$~$\rm km~s^{-1}~Mpc^{-1}$ and $\Omega_m = 1 - \Omega_{\Lambda} = 0.27$ for a flat universe. 
At the redshift of Abell~1775 ($z = 0.0717$), these parameters yield an angular diameter distance of 282.3~Mpc (i.e., $1\arcmin$ corresponds to 81~kpc) and a luminosity distance of 324.2~Mpc. We use the solar abundance standards of \citet{GS98}, according to which the iron abundance relative to hydrogen is $3.16\times10^{-5}$ in number.

%=====================================================================================

% Table 1
\begin{deluxetable}{cccccc}
\tablecolumns{6}
\tablecaption{Basic properties of Abell~1775 and two tailed radio galaxies. \label{tbl-1}}
\tablehead{
\colhead{Source Name}  &  \colhead{RA}  &  \colhead{Dec.}  &  \colhead{Redshift}  &  \colhead{Velocity\tablenotemark{a}}  &\colhead{$S_{235~\rm MHz}$\tablenotemark{b}} \\
\colhead{}  &  \colhead{(J2000)}  &  \colhead{(J2000)}  &  \colhead{$z$}  &  \colhead{($\rm km~s^{-1}$) }  &\colhead{(mJy)}}
\startdata
Abell~1775       &    $\rm 13h41m53.8s$       &     $\rm +26d22m19.0s$     &   0.0717   &  22551   &  ---   \\
			UGC~08669~NED01 (WAT)  &    $\rm 13h41m49.1s$       &     $\rm +26d22m24.5s$     &   0.0757   &  22704    &  135    \\
			UGC~08669~NED02 (NAT)  &    $\rm 13h41m50.5s$       &     $\rm +26d22m13.0s$     &   0.0694    &  20812    &  1900   \\
\enddata
\tablenotetext{a}{Radial velocity provided by \citet{zhang11}.}
\tablenotetext{b}{Flux density at 235~GHz measured with the GMRT \citep{giacintucci07}.}
\end{deluxetable}

%---------------------------------------------------------------------------------------------------
\clearpage

%####################################################################
\section{X-RAY OBSERVATION AND DATA REDUCTION}
%####################################################################
%##############################################
\subsection{ X-ray Observations}
%##############################################
%%##############################################
\subsubsection{\Chandra\ Data }
%##############################################
The \Chandra\ X-ray data analyzed in this work was obtained during two pointing observations performed on July 31, 2011 (ObsID 12891, 39.52 ks) and August 12, 2011 (ObsID 13510, 59.26 ks) with the S3 chip of the Advanced CCD Imaging Spectrometer (ACIS) operating in the VFAINT mode. In data reduction, we followed the standard steps suggested by the \Chandra\ X-ray Center by employing \Chandra\ Interactive Analysis of Observations (CIAO) v4.6 and \Chandra\ Calibration Database (CALDB) v4.7.0, as we did in our previous works \citep[e.g.,][]{hu19}. In brief, we adopted the CIAO script \texttt{chandra$\_$repro} to carry out corrections for gain, CTI, and astrometry, remove events with \ASCA\ grades 1, 5, and 7, and eliminate bad pixels and columns. To identify and exclude time intervals contaminated by occasional particle background flares, during which the background count rate deviates from the mean quiescent value by $20\%$, we examined the $0.5-12.0$~keV light curves extracted from the regions near the CCD edges, where the emissions from Abell~1775 and background sources are minor. Finally, we applied CIAO tools \texttt{wavdetect} and \texttt{celldetect} to identify and exclude all the point sources detected beyond the $3\sigma $ threshold in the ACIS images. These steps yielded a total of 98~ks clean data that will be studied in the following sections. 

%==================================
\subsubsection{\XMM\ Data}
%==================================
We have also analyzed the data obtained with the European Photon Imaging Camera (EPIC) onboard the \XMM\ observatory on February 20, 2004 (ObsID 0108460101, 33~ks), when all the detectors were set in Full Frame Mode with the MEDIUM filter. We followed the standard procedure \citep{snowden08} to carry out data reduction by using the Science Analysis System (SAS) v14.0.0. First, we applied the \XMM\ Extended Source Analysis Software (XMM-ESAS) tasks \texttt{emchain} and \texttt{epchain} to generate calibrated MOS and PN event files from raw data, respectively. In the screening process we set FLAG = 0 and kept events with PATTERNs $0-12$ for the MOS detectors and events with PATTERNs $0-4$ for the PN detector. Then, by examining the light curves extracted from source free regions in $1.0-5.0$~keV and $10.0-14.0$~keV, we rejected time intervals affected by soft and hard band flares, during which the detector count rate exceeds the $2\sigma$ limit above the quiescent mean value \citep[e.g.,][]{gu12,hu19}. We also identified and removed point sources by applying the tasks \texttt{cheese} and \texttt{cheese-bands}, and cross-checked the results by comparing them with those obtained with the \Chandra\ ACIS images (\S2.1.1).

%##############################################
\subsection{Background Templates}
%##############################################
We constructed the \Chandra\ and \XMM\ background templates by carrying out direct spectral fitting of the local background spectra obtained in the observations \citep[c.f.,][]{gu12,hu19}. Since the method has been described in depth in our recent work on the HCG~62 galaxy group \citep{hu19}, here we present a brief outline of the procedure of background modeling with an emphasis on the modeling of the particle induced background component.

The local background spectra were extracted from the boundary regions of the detectors, which are located at about $4\arcmin - 5\arcmin$ ($\sim 0.32-0.41$~Mpc) and $11\arcmin - 14\arcmin$ ($\sim 0.89-1.13$~Mpc) away from the cluster center for \Chandra\ and \XMM\ data, respectively. In these regions the ICM emission of Abell~1775 is relatively weak but cannot be ignored. Thus we approximated it by adopting an absorbed thermal APEC model, for which the absorption column density $N_{\rm H}$ was fixed at the Galactic value ($1.05 \times 10^{20}$~$\rm cm^{-2}$; \citealt{DL90}), while the gas temperature $kT$ and metal abundance $Z$ were set to be free. 
The background emission consists of three major independent components, i.e., the Galactic soft X-ray emission, the Cosmic X-ray Background, and the                                                                                                        Non-X-ray Background  \citep[e.g.,][]{gu12,hu19}, which were estimated as follows. 

\begin{enumerate}
\item \textbf{Galactic soft X-ray emission ---} This background component was modeled with one unabsorbed APEC component and one absorbed APEC component, which are assumed to be from the Local Hot Bubble and the Galactic Halo, respectively. Their temperatures were fixed at 0.08~keV and 0.2~keV, respectively, and their metal abundances were both set to 1 $\rm Z_{\sun}$ \citep[e.g.,][]{KS00,urban11}.

\item \textbf{Cosmic X-ray Background (CXB) ---} The CXB is dominated by the unresolved cosmic X-ray point source (i.e., AGN) and was modeled by using an absorbed power-law component with a photon index $\Gamma=1.4$ ($N_{\rm H}$ = $1.05 \times 10^{20}$~$\rm cm^{-2}$; see, e.g., \citealt{mushotzky00, CR07, gu12}).

\item \textbf{Non-X-ray Background (NXB) ---} The NXB component is mainly induced by the cosmic charged particles that interact with the detector and was modeled in different ways for \Chandra\ and \XMM. 
For \Chandra\ ACIS we employed the \Chandra\ stowed background data sets extracted from the same CCDs regions as used in this work to model the instrumental background. Owing to the stable detector background of the \Chandra\ ACIS, the normalizations of the stowed backgrounds simply scale with the count rates of the observation data in $9.5-12$~keV, where the \Chandra\ effective area is nearly zero \citep[e.g.,][]{vikhlinin05,HM06,sun09}. Meanwhile, a 3\% uncertainty was adopted on the normalization of the instrumental background to represent the systematic uncertainty in the current model \citep{vikhlinin05,sun09}.

For \XMM\ the NXB is further divided into two parts, i.e., the hard particle background (HPB) and the soft-protons background (SPB; \citealt{mernier15,hu19}). Since the quiescent particle background (QPB) makes the main contribution to the HPB, the HPB was modeled using the spectra extracted from the \XMM\ filter wheel closed (FWC) data by employing the \texttt{mos\_back} and \texttt{pn\_back} tools for the MOS and PN detectors, respectively. When modeling the HPB component, several Gaussian lines with free normalizations were also added into the QPB model (two Gaussians for the MOS spectra and six Gaussians for the PN spectra), because the QPB spectral regions that affected by the strong instrumental lines (e.g., Al K$\alpha$ and Si K$\alpha$ lines lying within $1.2-2.0$~keV for MOS) were cut out and replaced by an interpolated power-law (for details, see \citealt{snowden08}). On the other hand, although light curve filtering has been performed in our data reduction process (see \S2.1.2) as a primary treatment of the SPB, residual SPB contamination may still exist (see also \citealt{snowden08}). To account for the SPB residuals, we employed an unfolded broken power-law model with the $\Gamma$ parameter varying freely within $2.5-3.5$.
\end{enumerate}

Using the above models to fit the local background spectra, we obtained the best-fit background models and used them to create the corresponding background templates for each \Chandra\ and \XMM\ observation.

%###################
\section{RESULTS}
%###################

%====================================
\subsection{\Chandra\ and \XMM\ Imaging Analysis} 
%====================================

%==================================================
\subsubsection{General View} 
%==================================================

\begin{figure}
%\epsscale{1.0}
%\graphicspath{{figure/}}
\centering
\plotone{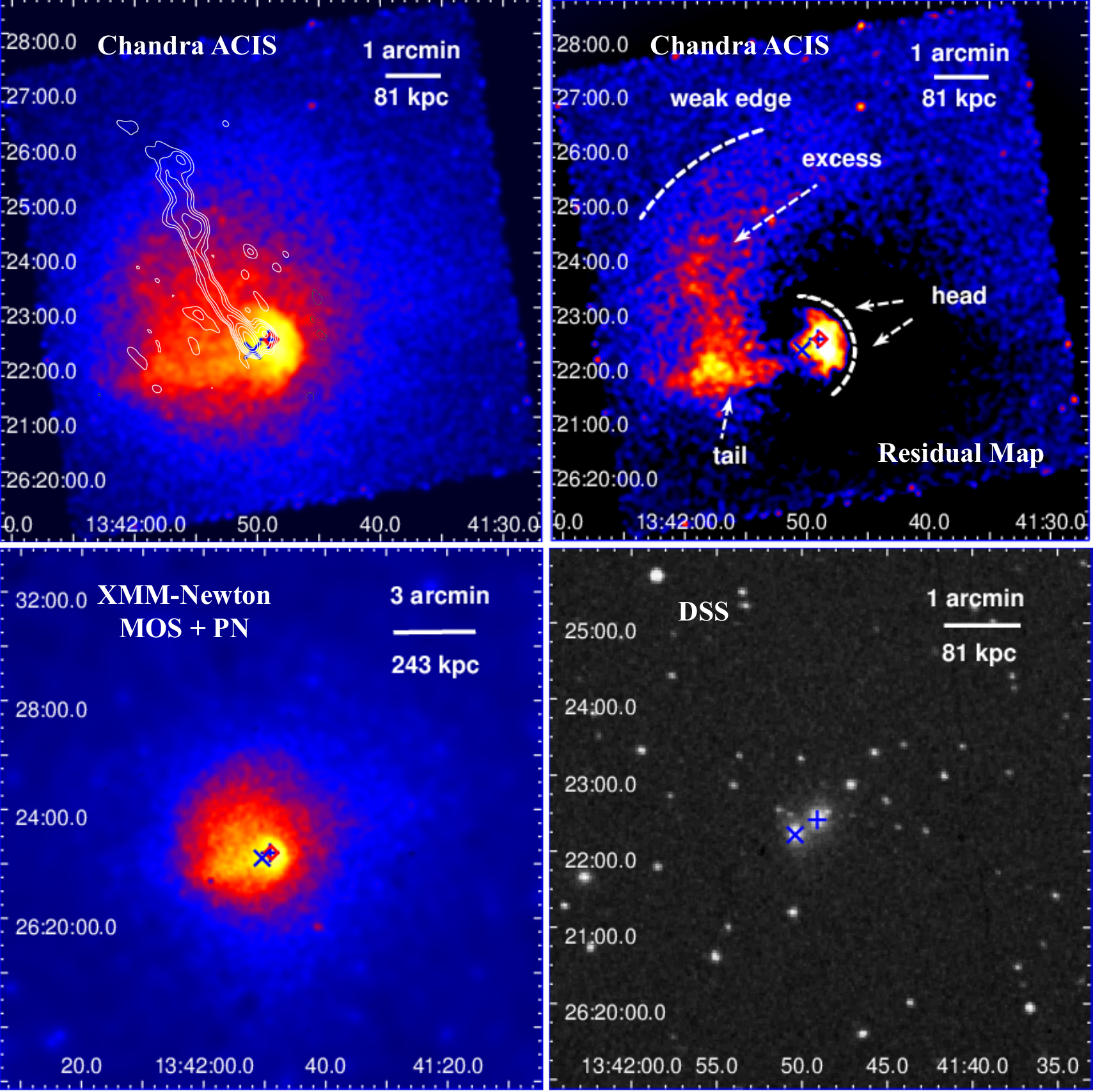}
\caption{Left: $0.5-7$~keV images of Abell~1775 obtained from two \Chandra\ observations (ACIS; top) \textbf{with 235~MHz GMRT radio contour from \citet{giacintucci07}} and one \XMM\ observation (MOS $+$ PN; bottom). Both images have been exposure-corrected and smoothed with a Gaussian kernel of \textbf{$1.5\arcsec$}, and are plotted in an asinh scale. Right: Residual map created by subtracting the best-fit 2D elliptical beta model from the \Chandra\ ACIS image (top) and the Digital Sky Survey (DSS) optical image (bottom).
The positions of the X-ray emission peak and the cores of the WAT and NAT radio galaxies are marked with `$\lozenge$', `$+$', and `$\times$', respectively.  \label{fig1}}
\end{figure}

\paragraph{\textbf{\Chandra\ ACIS Image}} \mbox{} \\
\indent The exposure-corrected $0.5-7$~keV ACIS image produced by using the data of the two \Chandra\ observations and the CIAO tool \texttt{merge$\_$obs} is shown in Figure~\ref{fig1}, in which the X-ray peak ($\rm R.A.=13^{h}41^{m}48.9^{s}, decl.=+26^{\degr}22^{\arcmin}25^{\arcsec}.5$, J2000.0) and the locations of the WAT and NAT tailed radio galaxies are marked with `$\lozenge$', `$+$' and `$\times$', respectively.
The X-ray emission of the cluster shows a clear ``mushroom-like'' morphology, including a wide arc-shaped edge (labeled as ``head'') at about 48~kpc west of the X-ray peak, which extends from southwest to north, and an obvious gas tail (labeled as ``tail''), which spans over $160$~kpc roughly toward the east. Since this type of head-tail feature is often seen in the regions where a dense (and usually cold) gas clump is undergoing ram pressure stripping caused by the relative motion between the gas clump and the environment \citep[e.g.,][]{russell10,million11,russell14}, it is reasonable to speculate that the cluster core of Abell~1775 is moving toward the northwest and the gas tail is stripped by ram pressure. 

%------------------------------------------------------------------------------------
% Figure 1
%------------------------------------------------------------------------------------

In order to better characterize these imaging features and search for possible fainter substructures, we employed \texttt{Sherpa} \citep{freeman01}, a modeling and fitting package embedded in CIAO, to fit the ACIS image with the two-dimensional (2D) elliptical beta model ({\it beta2D}). By using Cash statistic and Monte Carlo optimization method to constrain the fitting, we obtained the best-fit parameters as follows: core radius $r_{\rm 0} = 40$~kpc, ellipticity $ellip=0.02$, roll angle $theta=0.06$ radians, and slope of the profile $alpha=0.72$. After subtracting the best-fit 2D elliptical beta model, we obtained a residual map as shown in Figure~\ref{fig1}. 
In addition to the head and the tail, a plume of prominent X-ray emission excess (labeled as ``excess'') with an opening angle of $\sim 90\degr$ is revealed and is located at about $1\arcmin - 4\arcmin$ ($\sim 81-324$~kpc) northeast of the X-ray peak. The plume of emission excess shows a diffuse spiral pattern with a possible weak edge (labeled as ``weak edge'') at the northeast and almost vertically connects with the end of the X-ray tail. 
Since the X-ray excess may be related to the nearby host galaxy of the NAT, we have inspected the 300~MHz VLA image provided in \citet{GF00}. We find that only the top (i.e., the northwestern part) of the spiral excess partially overlaps the extremely long radio tail of the NAT radio galaxy, while the most part of the excess lies east of the NAT's radio tail.
This confirms the result of \citet{bliton98}, who briefly reported an X-ray enhancement to the northeast of the X-ray center by examining the \ROSAT\ PSPC map of Abell~1775 and found that the radio tail of the NAT coincides roughly with the X-ray enhancement. To explain this phenomenon, the authors qualitatively proposed that part of the interstellar medium stripped from the NAT galaxy has been compressed and adiabatic heated by ICM's bulk motion, thus the X-ray emission is enhanced locally. However, except that the plume of  emission excess as a whole is located at the northeast of the NAT's radio tail and only a small overlap between them, it is also interesting to see that the linear extension of the X-ray plume ($\sim 300$~kpc) is much larger than the width of the NAT's radio tail ($\sim 40$~kpc) at the junction, which is difficult to be explained with \citet{bliton98}. We will discuss the possible origin of this X-ray excess in \S3.2 and \S4.2.

\paragraph{\textbf{\XMM\ EPIC Image}}\mbox{} \\
\indent The background-subtracted and vignetting-corrected $0.5-7$~keV EPIC image (combined MOS and PN) produced by using the data of the \XMM\ observation and the SAS tasks \texttt{eimageget} and \texttt{eimagecombine} is also shown in Figure~\ref{fig1} for comparison. When generating the combined EPIC image, the task \texttt{eimageget} was used to create a set of images, including the observation image, scaled out-of-time (OOT; EPIC-PN only) image, scaled FWC image, vignetting corrected exposure maps, and mask image, for each EPIC detector (i.e., MOS1, MOS2, and PN) in the selected energy band. Then the task \texttt{eimagecombine} was used to combine the individual image output from \texttt{eimageget} to create the final science image. We find that in the outer regions ($\gtrsim 300$~kpc from the X-ray peak) the X-ray morphology of the cluster is approximately symmetric, while in the inner regions it exhibits nearly the same anisotropies as that revealed in the \Chandra\ image (Fig.~\ref{fig1}).

%==================================================
\subsubsection{Possible AGNs Associated with the Two Tailed Radio Galaxies} 
%==================================================

In Figure~\ref{fig1}, we also mark the locations of the NAT and WAT radio galaxies in the $0.12\degr \times 0.12\degr$ Digital Sky Survey (DSS) image. We find that the position of the BCG, the host of the WAT, coincides with the X-ray peak within $\simeq 3\arcsec$ (or 4~kpc). The BCG contains an X-ray point source in its core, which can be easily detected by applying the CIAO tool \texttt{wavdetect} (with scales of 1, 2, 4, 8, and 16 pixel and a detection threshold of $10^{-6}$) in the $2-10$~keV \Chandra\ ACIS image. In order to investigate the nature of this point source, we extracted the spectrum from a circular region with a radius of $4\arcsec$, which is centered at the point source, subtracted the contaminating cluster emission from it by utilizing the spectrum extracted from a neighboring annular region that covers $10\arcsec-50\arcsec$ from the point source \citep[e.g.,][]{terashima03}, and fitted the obtained spectrum with an absorbed power-law model. We find that the best-fit photon index is $\Gamma = 1.76 \pm 0.18$, and the corresponding $2-10$~keV luminosity is $L_{2-10~\rm keV} = 2.56\times 10^{41}$~$\rm erg~s^{-1}$, which are both consistent with those of a typical low luminosity AGN (LLAGN; $L_{\rm X} < 10^{42}$~$\rm erg~s^{-1}$; \citealt{ho01,terashima03}). From the \Chandra\ image and the residual map, however, we have not found any clear evidence for X-ray cavities around the BCG, possibly because the morphology of the cluster core is asymmetric or the putative X-ray cavity is small and its significance is reduced by the projection effect \citep{shin16}.
As for the NAT radio galaxy, no X-ray point source can be detected using \texttt{wavdetect}; based on the \ROSAT\ PSPC data, \citet{bliton98} also failed to identify any X-ray point source at the position of the NAT with an upper limit of $L_{2-10~\rm keV}$ $< 0.6\times 10^{41}$~$\rm erg~s^{-1}$.

%==================================================
\subsubsection{KHI Features on the Cold Front and the Stripped Tail of the Moving Gas Core} 
%==================================================

\begin{figure}
%\graphicspath{{figure/}}
\centering
%\plotone{fig2-radio.pdf}
\includegraphics[scale=.35]{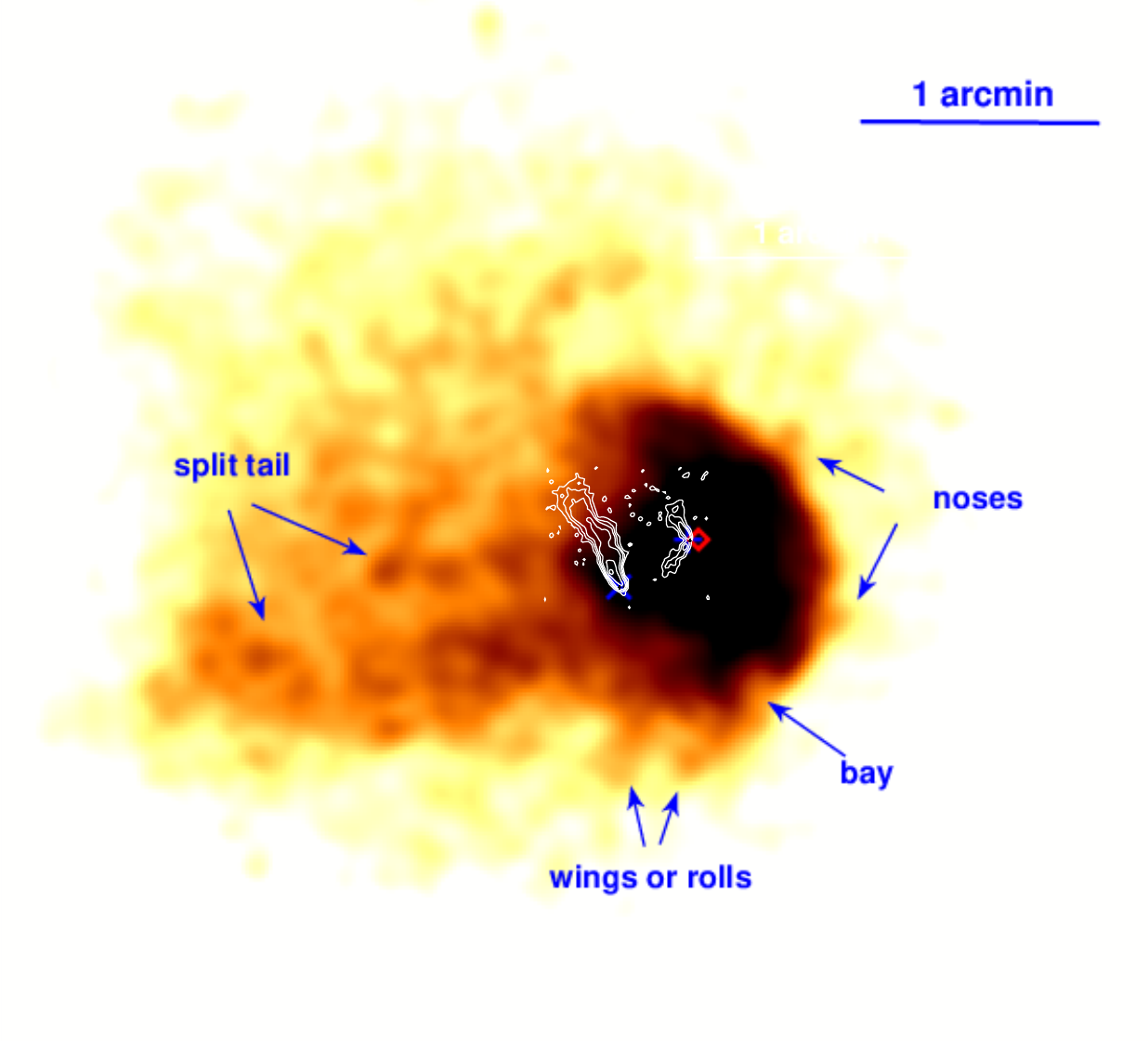}
\includegraphics[scale=.30]{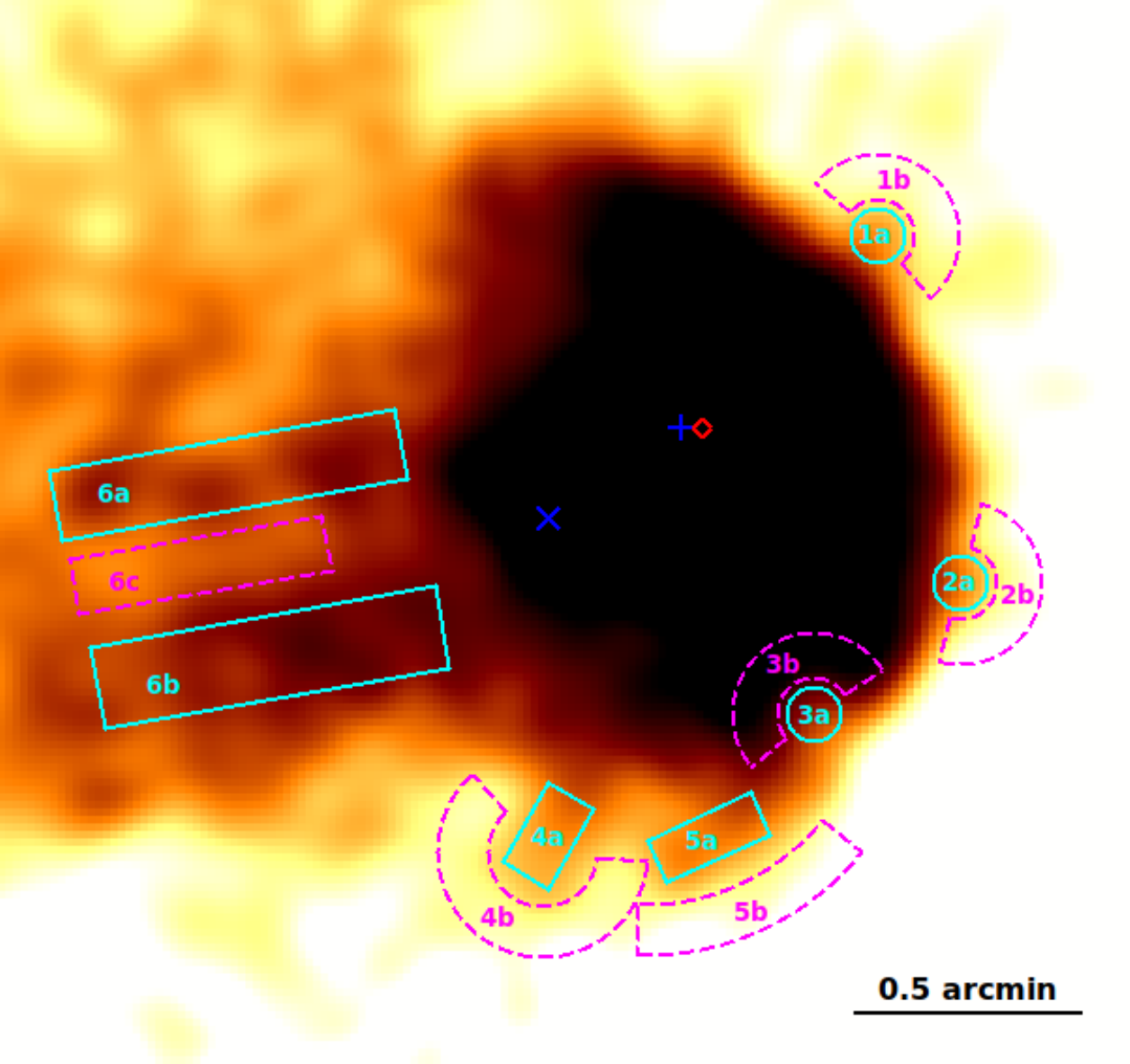}
\caption{Left: Zoom-in view of the central $4\arcmin \times 4\arcmin$ of the $0.5-7$~keV \Chandra\ ACIS image \textbf{smoothed with a Gaussian kernel of 3\arcsec. The 1.45~GHz VLA radio contours are overlaid}. The characteristic KHI features, i.e., noses, concave bay, wings (or rolls), and the split tail, are indicated by arrows. \textbf{Right: Zoom-in view of the left figure. The regions with cyan color cover the KHI features, and the adjacent regions used for comparison are also highlighted with magenta color.} \label{fig2}}
\end{figure}

In order to better examine the substructures in the inner regions, we enlarge the central $4\arcmin \times 4\arcmin$ of the $0.5-7$~keV \Chandra\ ACIS image and plot in Figure~\ref{fig2}. Interesting features usually attributed to the Kelvin-Helmholtz instabilities (KHIs), which is caused by the shear flow along the interface between two kinds of fluids, can be easily identified at the boundary of the moving gas core. These include two small but apparent noses on the head, two wings (or rolls) at the south, one striking concave bay at the southwest, and a split gas tail, which are all labeled in Figure~\ref{fig2}. 
Among these, the noses and wings (or rolls) define two common forms of distortion generated by the shear flow along the cold front, and have been detected in the galaxy groups NGC~7618 and UGC~12491 \citep{roediger12b}.
The origin of the concave bay can be explained by assuming that two KHI rolls are developing on both sides of it, as indicated in some simulations \citep[e.g.,][]{roediger12a,roediger13}. Except in the core region, such concave bays have also been observed on the outer cold fronts, whose radii reach tens of kpc, in Perseus, Centaurus, and Abell 1795 \citep{walker17}, which are also associated with KHI rolls that are larger than their counterparts in cluster core regions by a factor of $\sim 10$. The appearance of the split tail, the morphology of which is similar to that found in the UGC~12491 group \citep{roediger12b}, on the other hand, can also be interpreted as the direct result of the KHI distortions caused by the velocity shear. 

\textbf{To examine the significances of these KHI features, we have calculated the photon counts for each of the KHI features and their adjacent ambient medium \citep{roediger12b} and adopted the u-test \citep{CL66} since the photon counts obey Poisson distribution. For two regions with areas S1 and S2 that have photon counts x and y, respectively, u value \citep{sichel73} can be defined as
\begin{equation}
u=\frac{cX-Y}{(c^{2}X+Y)^{1/2}} ,
\end{equation}
where c (= S2/S1) is the ratio of areas. Assuming x and y are the means of the Poisson distributions X and Y, respectively, u value is supposed to obey $N(0,1)$ if X is identical to Y. As shown in Figure~\ref{fig2}, we have selected seven regions over the KHI features marked with cyan and six adjacent regions that use for comparison marked with magenta. The photon counts and the areas of all regions, as well as the u values of the KHI features are listed in Table~\ref{tbl-2}. We find that u values of the KHI features imply a significant level of at least $3\sigma$.}
%We have selected 7 regions over the KHI features, which are plotted in Figure~\ref{fig2} with cyan color. Correspondingly, the adjacent regions used for comparison are also highlighted in Figure~\ref{fig2} with magenta color. }
%The number of counts and areas in each region are listed in Table . 
%We find that all the KHI features are significant at least $2.9\sigma$ above the adjacent medium.}
%------------------------------------------------------------------------------------

% Figure 2

%-----------------------------------------------------------------------------------

% Table 2
\begin{deluxetable}{cccccccccccccc}
%\tablewidth{0.pt}
\centering
\tablecolumns{14}
\tablecaption{\textbf{The photon counts and the areas of all regions shown in Figure~\ref{fig2}, as well as the u values of the KHI features.} \label{tbl-2}}
\tablehead{
\colhead{}  &  \multicolumn{2}{c}{nose}  &  \multicolumn{2}{c}{nose}  & \multicolumn{2}{c}{bay}   & \multicolumn{2}{c}{wing} &   \multicolumn{2}{c}{wing}  & \multicolumn{3}{c}{tail}\\
\colhead{Region}  &  \colhead{1a}  &  \colhead{1b}  &  \colhead{2a}  &  \colhead{2b}  &  \colhead{3a}   &  \colhead{3b} &  \colhead{4a}   &  \colhead{4b} &  \colhead{5a}   &  \colhead{5b} &  \colhead{6a}   &  \colhead{6b}  &  \colhead{6c}}
\startdata
Counts   & 119 &  279   &  119   &  256   & 121 & 689 & 204 & 489 & 258 & 339 & 1328 & 1719  &  671  \\
\hline

Area & \multirow{2}{*}{40.7}  &   \multirow{2}{*}{153}  &   \multirow{2}{*}{39.7}  &  \multirow{2}{*}{145}   &  \multirow{2}{*}{39.7}  &  \multirow{2}{*}{152} &  \multirow{2}{*}{84} &  \multirow{2}{*}{269}  &  \multirow{2}{*}{90}  &  \multirow{2}{*}{206} &  \multirow{2}{*}{421}  & \multirow{2}{*}{502} &  \multirow{2}{*}{244}  \\
($\rm arcsec^{2}$)  &    &      &    &   &    &    &   &    &    &    &     &   &     \\
\hline
u-statistic &  3.8$\sigma$ &  & 4.2$\sigma$ &   &  4.5$\sigma$ &   & 3.2$\sigma$ &   &  6.1$\sigma$ &   & 3.0$\sigma$  & 5.0$\sigma$  &   \\
\enddata
\end{deluxetable}
%--------------------------------------------------------------------------

%-----------------------------------------------
\subsection{\Chandra\ Imaging and Spectral Analysis in the Three Sectors}
%-----------------------------------------------

\begin{figure}
%\graphicspath{{figure/}}
\centering
\includegraphics[scale=.3]{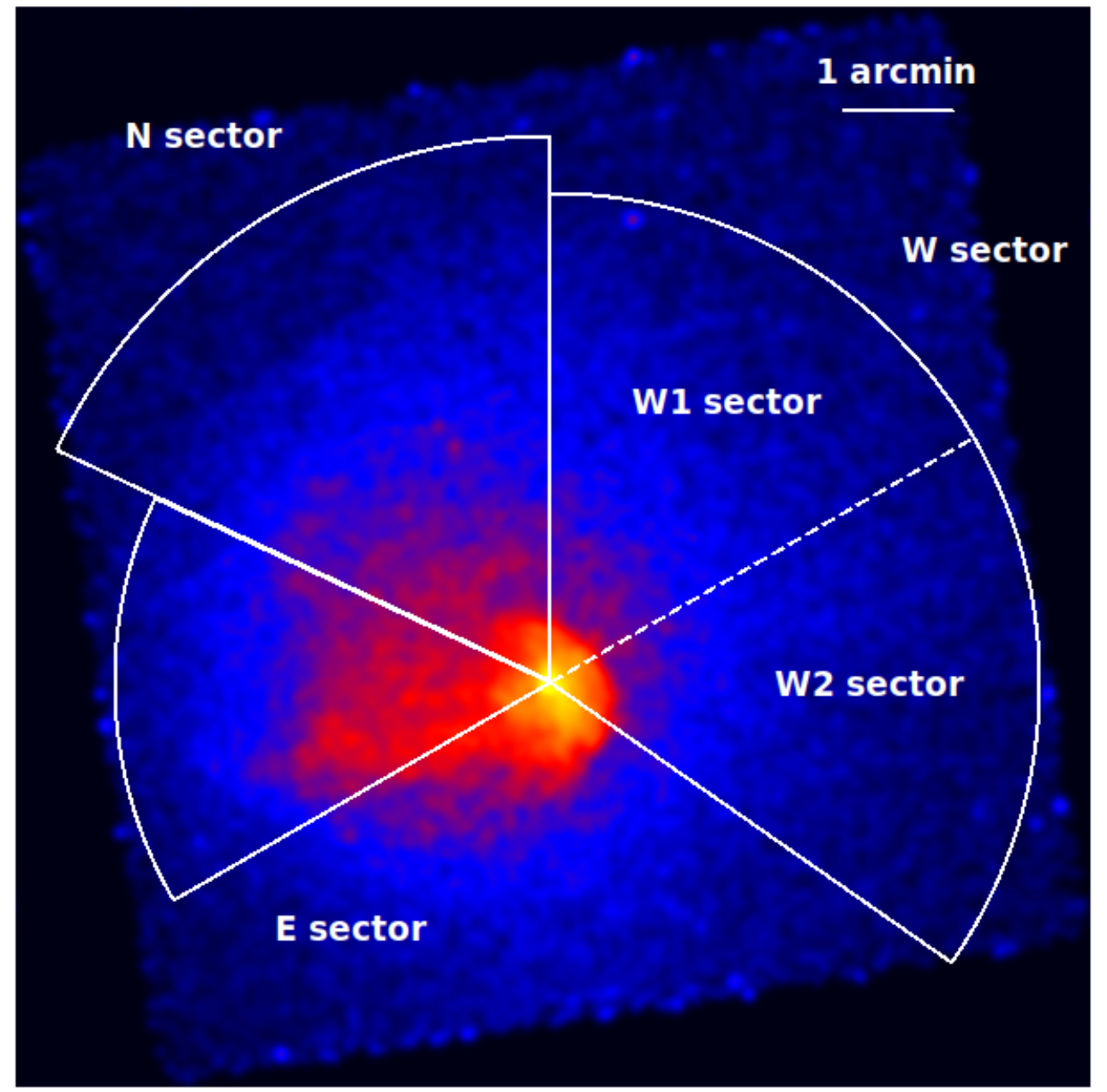}
\includegraphics[scale=.5]{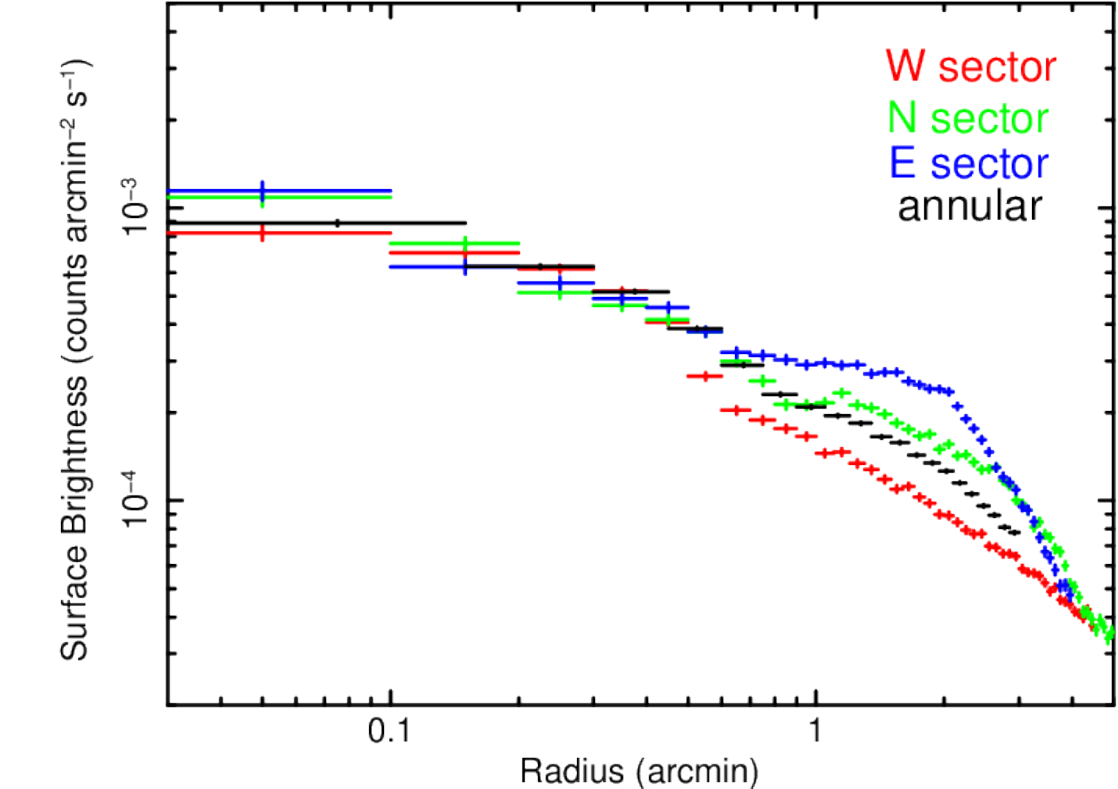}
\caption{Left: Sectors used to extract radial profiles of the X-ray surface brightness and gas temperature. Right: X-ray surface brightness profiles extracted from the three sectors together with the azimuthally averaged X-ray surface brightness profile extracted in a set of concentric annuli.  \label{fig3}}
\end{figure}

In order to clarify the origins of the detected X-ray substructures, we study the gas properties in a quantitative way by analyzing the \Chandra\ X-ray surface brightness profiles and the X-ray spectra extracted in $0.5-7.0$~keV from three pie-region sets defined in three circular sectors (i.e., W sector, E sector, and N sector shown in Figure~\ref{fig3}), which roughly cover the head, the tail and the region exhibiting the X-ray emission excess, respectively. \textbf{W sector is further divided into W1 and W2 sectors with equal angles base on the residue map (Fig.~\ref{fig1}), since the gas resided in the W1 sector might be mixed by the sloshing gas, while the gas in the W2 sector is more likely to be unpolluted ambient gas.}
%-------------------------------------------------------------------------------
% Figure 3
%-------------------------------------------------------------------------------
The X-ray surface brightness profiles were extracted from the merged \Chandra\ ACIS image, whereas, the X-ray spectra were extracted from the two \Chandra\ observations separately, which will be studied jointly in the spectral fittings.
In the spectral analysis, we created the appropriate Ancillary Response Files (ARFs) and Redistribution Matrix Files (RMFs) by using the CIAO tool \texttt{specextract}, and fitted the spectra by using XSPEC version 12.8.2 \citep{arnaud96} and AtomDB v3.0.8.
To model the X-ray emission of the ICM, we employed the Astrophysical Plasma Emission Code (APEC) model, which is designed for a plasma in collision ionization equilibrium \citep{smith01}, and set both the gas temperature and metal abundance free. Since there is no evidence for an extra absorption, we adopted an absorption associated with Galactic column density ($N_{\rm H} = 3.00 \times 10^{20}$~$\rm cm^{-2}$; \citealt{DL90}). Considering that the X-ray morphology of the central part of Abell~1775 is far from spherical symmetry, we will follow, e.g., \citet{mernier15}, and focus on projected analysis. To model the spectra extracted in the central 0\arcmin.6, we have also attempted to add another APEC component to represent the possible contribution of the cooler phase gas \citep[e.g.,][]{hu19}. However, by applying the statistical $F$-test to distinguish between the single-temperature and two-temperature models, we obtained a probability of 0.31, which indicates that there is no significant improvement in the fittings after introducing the two-temperature model. Thus the second APEC component is ignored throughout this work. Since all the single-temperature fittings are acceptable here, we summarize the best-fit parameters together with their 90\% confidence ranges in Table~\ref{tbl-3}.

%-------------------------------------------------------------------------------

% Table 3
\begin{deluxetable}{cccc}
\tabletypesize{\scriptsize}
\tablewidth{0.pt}
\centering
\tablecolumns{4}
\tablecaption{Best-fit model parameters obtained in the projected spectral analysis for the W, E, and N sectors \label{tbl-3}}
\tablehead{
\colhead{Radius}  &  \colhead{kT}  &  \colhead{Z}  &  \colhead{$\chi^{2}$/dof} \\
\colhead{(arcmin)}  &  \colhead{(keV)}  &  \colhead{($Z_{\sun}$)}  & }
\startdata
\multicolumn{4}{l}{W sector} \\
\tableline
$0.0 - 0.6$ & $ 3.39_{-0.18}^{+0.28} $ & $ 0.93_{-0.19}^{+0.23}$ &  1656.41/1637(1.01) \\
$0.6 - 1.2$ & $ 5.30_{-0.43}^{+0.54} $ & $ 0.57_{-0.22}^{+0.25}$   \\
$1.2 - 2.0$ & $ 4.85 \pm 0.38 $ & $ 0.55_{-0.17}^{+0.19}$   \\
$2.0 - 3.0$ & $ 4.07_{-0.27}^{+0.28} $ & $ 0.45_{-0.14}^{+0.16}$  \\
$3.0 - 4.5$ & $ 3.50_{-0.19}^{+0.28} $ & $ 0.37_{-0.11}^{+0.12}$  \\
\tableline
\multicolumn{4}{l}{E sector} \\
\tableline 
$0.0 - 0.6$ & $ 4.03_{-0.40}^{+0.43} $ & $ 0.67_{-0.28}^{+0.37}$ & 1250.57/1209(1.03) \\
$0.6 - 1.2$ & $ 4.06_{-0.30}^{+0.31} $ & $ 0.68_{-0.20}^{+0.24}$  \\
$1.2 - 2.0$ & $ 3.41_{-0.14}^{+0.20} $ & $ 0.84_{-0.14}^{+0.15}$  \\
$2.0 - 3.0$ & $ 2.96 \pm 0.15 $ & $ 0.64_{-0.10}^{+0.12}$  \\
$3.0 - 4.0$ & $ 2.90_{-0.23}^{+0.24} $ & $ 0.51_{-0.14}^{+0.17}$  \\
\tableline 
\multicolumn{4}{l}{N sector} \\
\tableline 
$0.0 - 0.6$ & $ 4.23_{-0.34}^{+0.47} $ & $ 1.29_{-0.38}^{+0.47}$ & 1397.81/1353(1.03)  \\
$0.6 - 1.2$ & $ 4.93_{-0.48}^{+0.46} $ & $ 0.83_{-0.29}^{+0.35}$   \\
$1.2 - 2.0$ & $ 4.43_{-0.28}^{+0.37} $ & $ 0.64_{-0.18}^{+0.20}$  \\
$2.0 - 3.0$ & $ 3.32_{-0.16}^{+0.19} $ & $ 0.62_{-0.12}^{+0.14}$  \\
$3.0 - 4.0$ & $ 2.70_{-0.15}^{+0.20} $ & $ 0.48_{-0.10}^{+0.12}$  \\
$4.0 - 5.0$ & $ 3.01_{-0.33}^{+0.37} $ & $ 0.47_{-0.17}^{+0.24}$  \\
\tableline
\enddata
\end{deluxetable}

%-----------------------------------------------------------
\subsubsection{W Sector: A Cold Front Preceding the Cluster Core}
%-----------------------------------------------------------

%-----------------------------------------------------------------------------------
% Figure 4
\begin{figure}
\centering
\includegraphics[scale=.4]{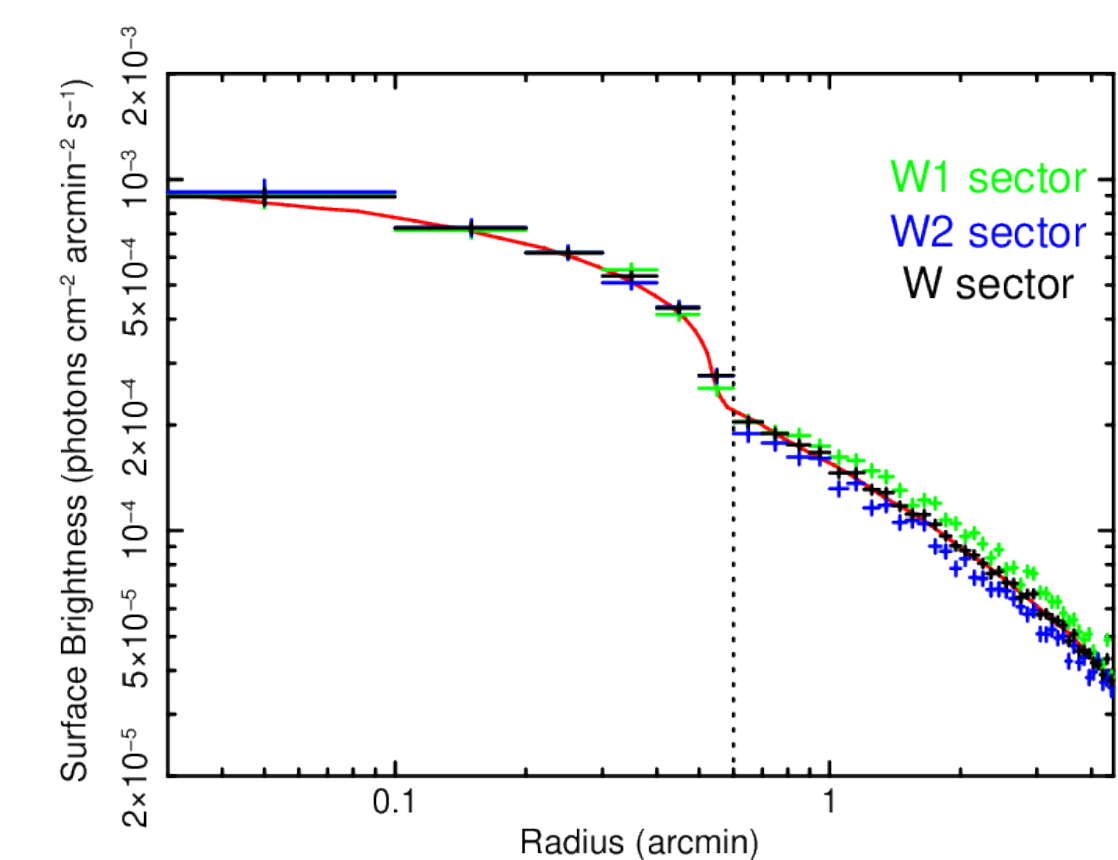}
\includegraphics[scale=.4]{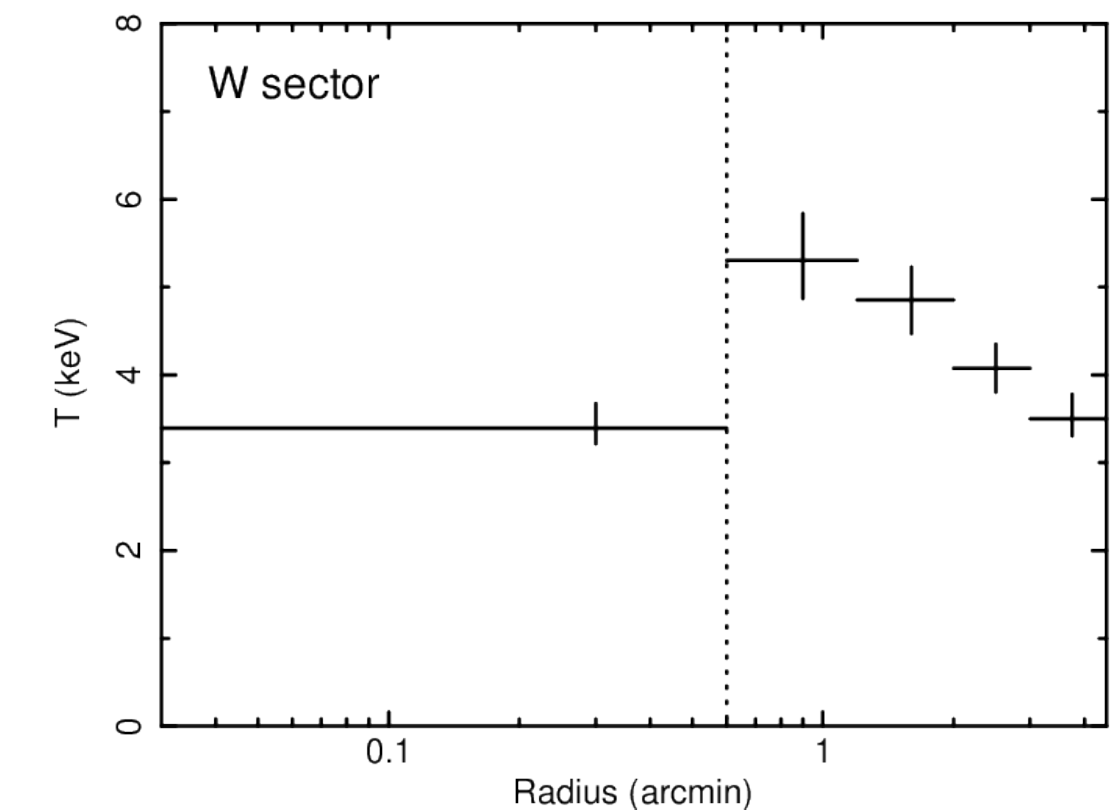}
\caption{Left: X-ray surface brightness profiles obtained in the W sector, W1 sector, and W2 sector. Right: The gas temperature profile derived in the W sector. The dotted line marks the location of the edge (i.e., head), and the red solid line represents the best-fit model of the X-ray surface brightness profile measured in the W sector, which is calculated by using a 3D broken power-law gas density model (\S3.2.1).  \label{fig4}}
\end{figure}
%-----------------------------------------------------------------------------------

We show spatial distributions of the X-ray surface brightness and projected gas temperature in the W sector in Figure~\ref{fig4}. The head revealed in the X-ray images (Fig.~\ref{fig1}) can be identified as a jump in both of the observed profiles at about $48$~kpc west of the X-ray centroid. To better characterize this emission discontinuity, we fit the derived X-ray surface brightness profile by applying the software PROFFIT v.1.5 \citep{eckert11,eckert16}, which projects the model predicted gas emission along the line of the sight based on the following three-dimensional (3D) density profile
\begin{equation}
n_{\rm e}(r) = 
\begin{cases}
C r^{-\alpha_{1}}  & \quad \rm{if}~r \leq r_{\rm break} \\
C \frac{1}{d_{\rm jump}} r^{-\alpha_{2}}  & \quad \rm{if}~r>r_{\rm break}
\end{cases},
\end{equation}
where $r_{\rm break}$ is the break radius, $\alpha_{1}$ and $\alpha_{2}$ are the inner and outer slope indices of the gas density profile, respectively, $d_{\rm jump}=n_{\rm e,in}/n_{\rm e,out}$ measures the degree of the gas density jump, and $C$ is the normalization.
The best-fit ($\chi_{\nu}^{2} = 0.80$; Fig.~\ref{fig4}) parameters are $\alpha_{1}=0.33\pm0.03$, $\alpha_{2}=0.75\pm0.01$, $r_{\rm break}=0\arcmin.56\pm0\arcmin.01$, and $d_{\rm jump}=1.80\pm0.02$. 
Meanwhile, at the position of the jump, the projected gas temperature rises from $3.39_{-0.18}^{+0.28}$~keV to $5.30_{-0.43}^{+0.54}$~keV outwards, which indicates that at the 90\% confidence level, there exists a cold front caused by the motion of a cold, dense gas clump in the ambient hot ICM.

In order to search for the possible shock often seen in the similar merging systems, which precedes the cold front, we have carefully examined the surface brightness profiles of W, W1, and W2 sectors via both visual inspection and model fitting with the same broken power-law gas density model as adopted above. However, no evidence is found for the existence of such a shock after testing with different region partitions. This is different from some well-known merging systems, such as Abell~3667 \citep{vikhlinin01}, 1E0657$-$56 (Bullet cluster; \citealt{markevitch02,markevitch06}), Abell~2146 \citep{russell11}, RX~J0751.3$+$5012 \citep{russell14}, Abell~665 \citep{dasadia16}, and MACS J0553.4$-$3342 \citep{botteon18}, in which a shock front is formed in front of the core of the infalling subcluster, with the cold front sitting roughly in the middle of them. 
Since both observations and simulations \citep[e.g.,][]{SF07,ML13,machado15,zhang19} show that shock fronts are more likely to be detected when the velocity of the infalling subcluster remarkably exceeds the sound speed in the ambient ICM, the absence of the shock front in Abell~1775 indicates that the cold gas core is likely to be moving at a subsonic speed. Further discussion on this will be presented in \S4.1.

%----------------------------------------------------------
\subsubsection{E Sector: A Gas Tail Caused by Ram Pressure Stripping}
\begin{figure}
\centering
\includegraphics[scale=.4]{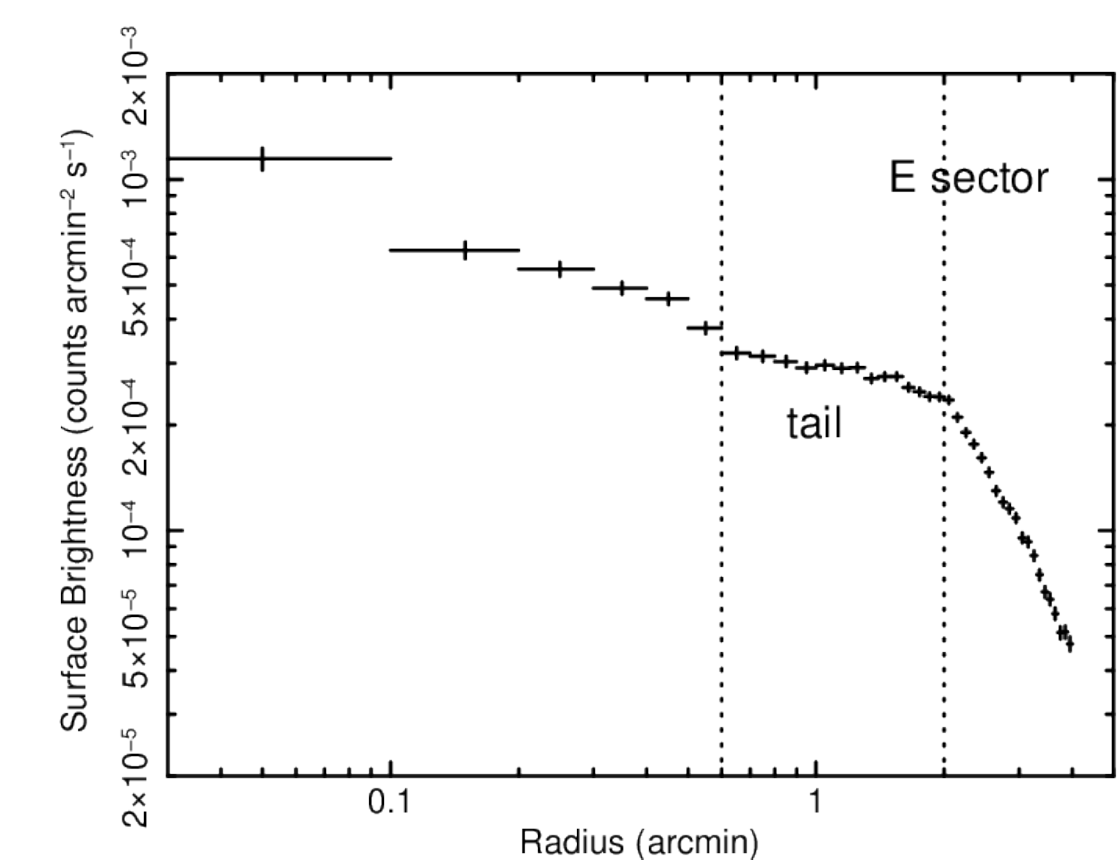}
\includegraphics[scale=.4]{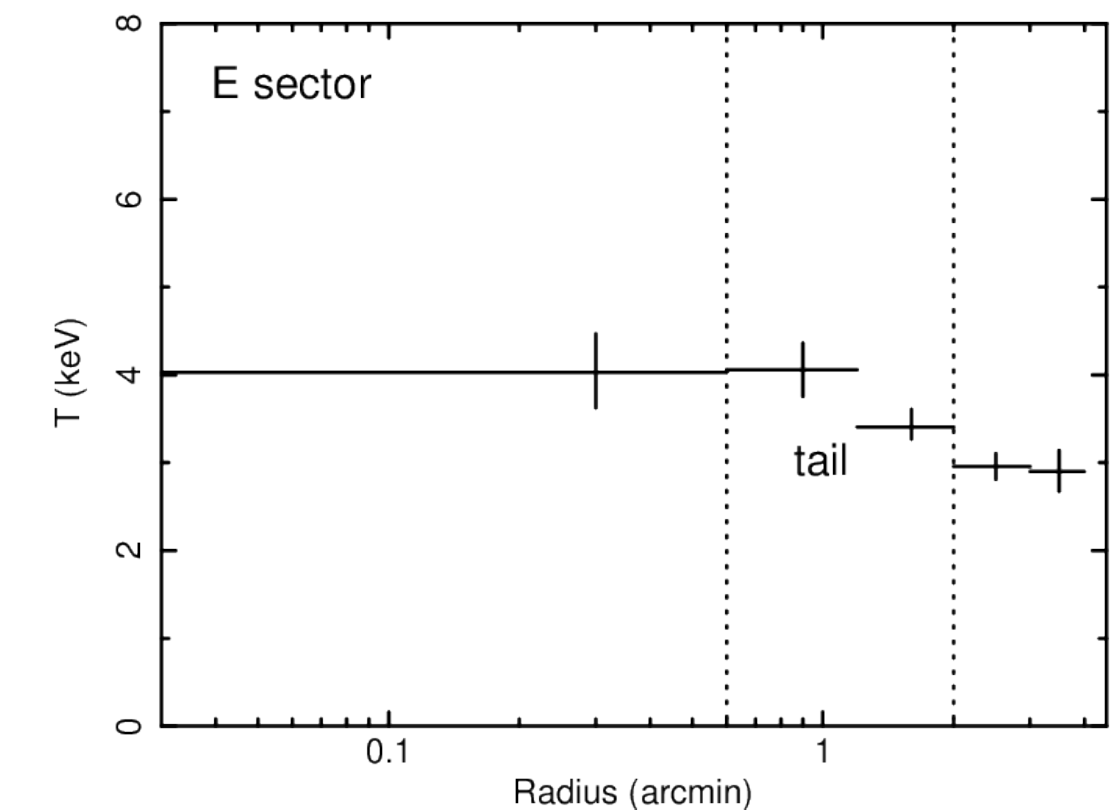}
\caption{X-ray surface brightness profile (left) and projected gas temperature profile measured in the E sector (right). \label{fig5}} 
\end{figure}

%----------------------------------------------------------
In Figure~\ref{fig5}, we show the radial distributions of the X-ray surface brightness and gas temperature along the gas tail, which extends to about 2\arcmin\ (i.e., 162~kpc) east of the X-ray peak. An evident X-ray emission plateau is revealed in about $0\arcmin.6 - 2\arcmin$, where the gas temperature declines slightly from $4.06_{-0.30}^{+0.31}$~keV to $2.96\pm0.15$~keV.
Since this temperature range is consistent with that of in the core of Abell~1775 ($4.03_{-0.40}^{+0.43}$~keV), the gas tail is very likely to have been formed by the gas expelled out of the cluster core due to the ram pressure stripping during the westward or northwestward motion of the core.
We will discuss the formation of the gas tail by calculating the core velocity relative to the ambient gas in \S4.1.

%-----------------------------------------------------------------------------------

% Figure 5

%-----------------------------------------------------------------------------------

%-------------------------------------------------------
\subsubsection{N Sector: A Spiral-like X-ray Emission Excess}
\begin{figure}
%\graphicspath{{figure/}}
\centering
\includegraphics[scale=.4]{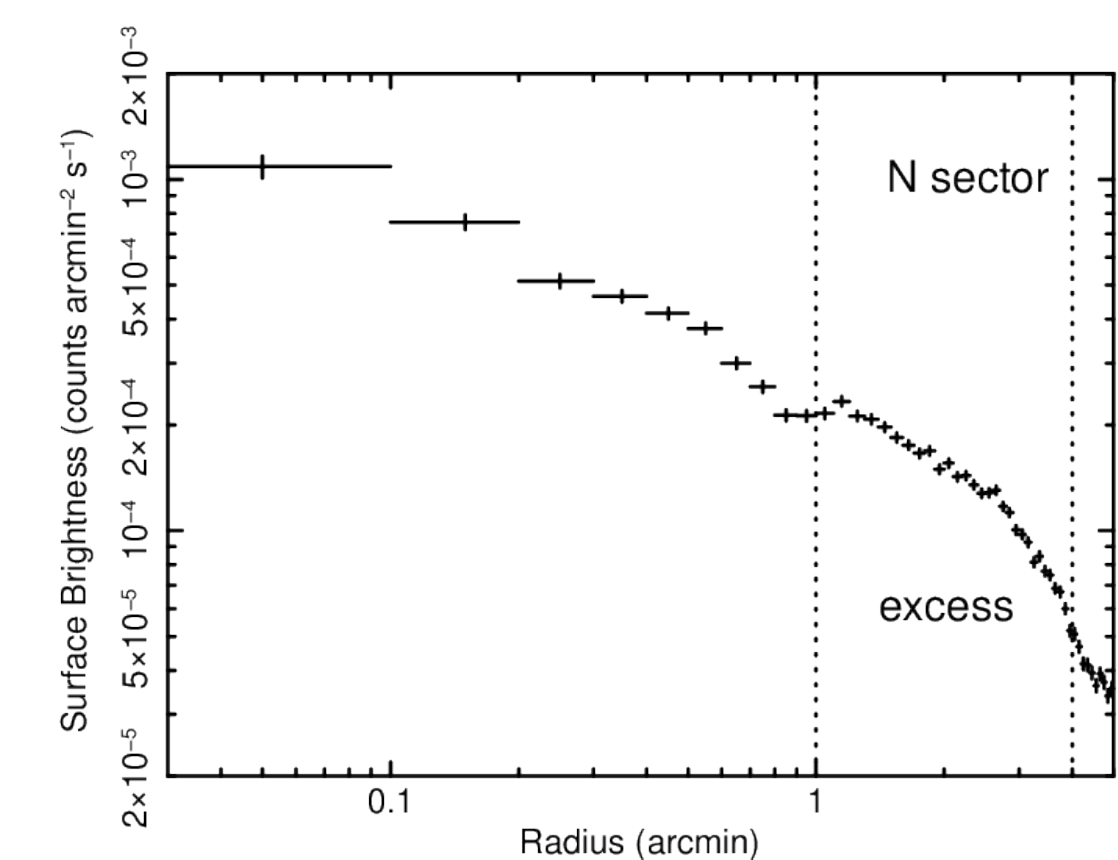}
\includegraphics[scale=.4]{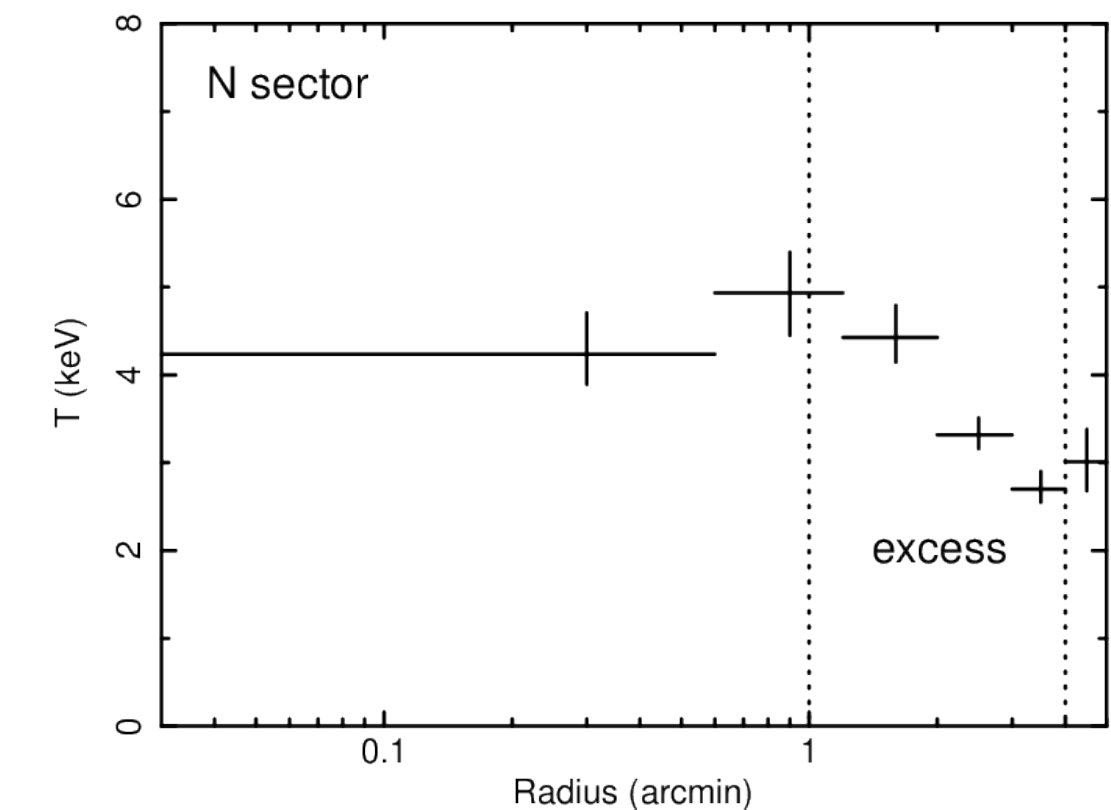}
\caption{X-ray surface brightness profile (left) and projected gas temperature profile measured in the N sector (right). \label{fig6}}
\end{figure}

%-------------------------------------------------------
We find that the X-ray surface brightness excess shown in Figures~\ref{fig1} and \ref{fig2} is located at about $1\arcmin - 4\arcmin$ (i.e., $81-324$~kpc) from the X-ray peak in the N sector (Figs.~\ref{fig3} and~\ref{fig6}). In this sector, the projected gas temperature rises mildly from $4.23_{-0.34}^{+0.47}$~keV in the innermost region ($<0\arcmin.6$) to $4.93_{-0.48}^{+0.46}$~keV in $0\arcmin.6 - 1\arcmin.2$, and then drops to $4.43_{-0.28}^{+0.37}$~keV in $1\arcmin.2-2\arcmin$ (i.e., the inner part of the excess). In $2\arcmin-4\arcmin$ (i.e., the outer part of the excess), however, the gas temperature decreases quickly to about 3~keV, which is comparable to the gas temperature at the end of the tail (about 2\arcmin\ in the E sector). 
Considering this similarity, as well as the spatial connection between the excess and the gas tail at its end (Fig.~\ref{fig1}), it is natural to speculate that the cold gas in the excess may have the same origin as that contained in the tail. This possibility will be discussed via numerical simulations in \S 4.2.

%------------------------------------------------------------------------------------
% Figure 6
%------------------------------------------------------------------------------------

We also studied the possible weak edge identified at the northeast of the emission excess region (see the residual map in Fig~\ref{fig1}) by fitting the surface brightness profile extracted in the N sector using the same gas density model described in \S3.2.1. We find that this possible weak edge is located at $4\arcmin.06 \pm0\arcmin.03$ from the X-ray peak and the density jump at the edge is only $1.18\pm0.09$. However, we failed to detect any temperature jump at this edge, possibly because the temperature jump, if it does exist, is marginal or mild.

%-----------------------------------------------
\subsection{\Chandra\ Two-Dimensional Thermodynamic Maps}
%-----------------------------------------------

In order to provide further insights into the overall thermal dynamical state of the X-ray gas halo in Abell~1775, we generated the 2D temperature and metal abundance maps based on the high-resolution \Chandra\ ACIS data by employing the following two approaches: (1) improved Centroidal Voronoi Tessellation (CVT; \citealt{CC03}); and (2) Contour Binning \citep{sanders06}. 
The CVT approach is an adaptive binning technique, which defines the seed of each Voronoi cell as the cell's centroid, and has been widely adopted in literatures \citep[e.g.,][]{gu07,randall08,gu12}. In this work, we first employed the CVT algorithm to divide the cluster into Voronoi cells with suitable sizes, so that each cell contains a total of approximately 200 photons from two observations. For each cell we further defined a larger circle centered at it, for which a total of about 2000 photons is guaranteed from two observations, and extracted its spectra. 
The second approach adopted in this work was developed by \citet{sanders06} and has been used to partition spatial bins basically based on the intensity contours generated from the adaptive smoothed surface brightness map. Each bin was created by combining adjacent pixels until that the signal-to-noise ratio reaches a threshold of 30.  
For both approaches, the spectra and response files were prepared for each bin. 
As indicated by \citet{sanders06}, we find that the contour binning method is well suited to tracing the spatial substructures (e.g., edges), that are more correlated with the surface brightness distribution, especially in the core regions, and the improved CVT method, on the other hand, has an advantage of being sensitive to reveal substructures that are uncorrelated with the surface brightness distribution \citep{osullivan19}. 

\begin{figure}
\epsscale{1}
%\graphicspath{{figure/}}
\centering
\includegraphics[scale=.4]{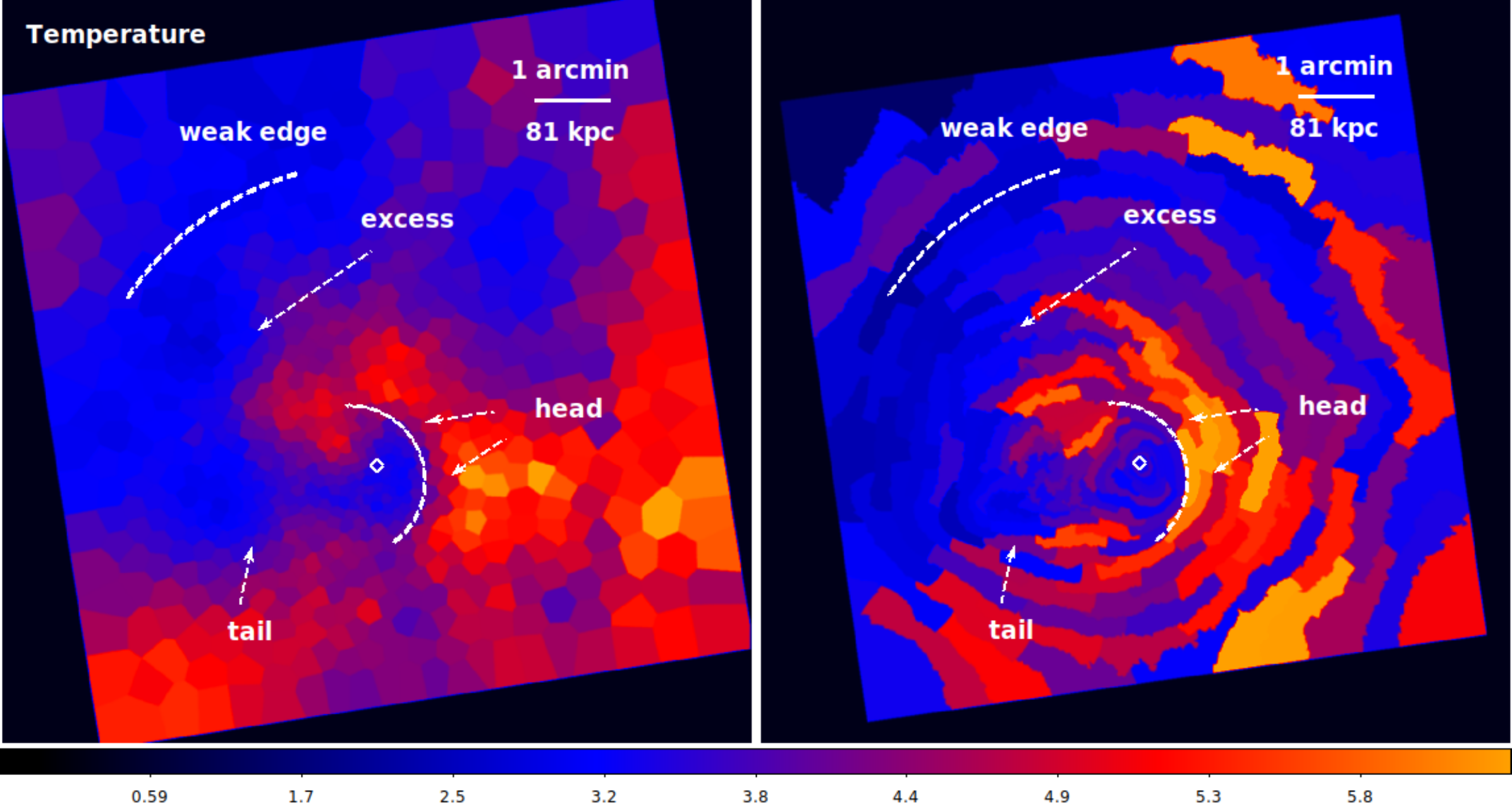}
\includegraphics[scale=.4]{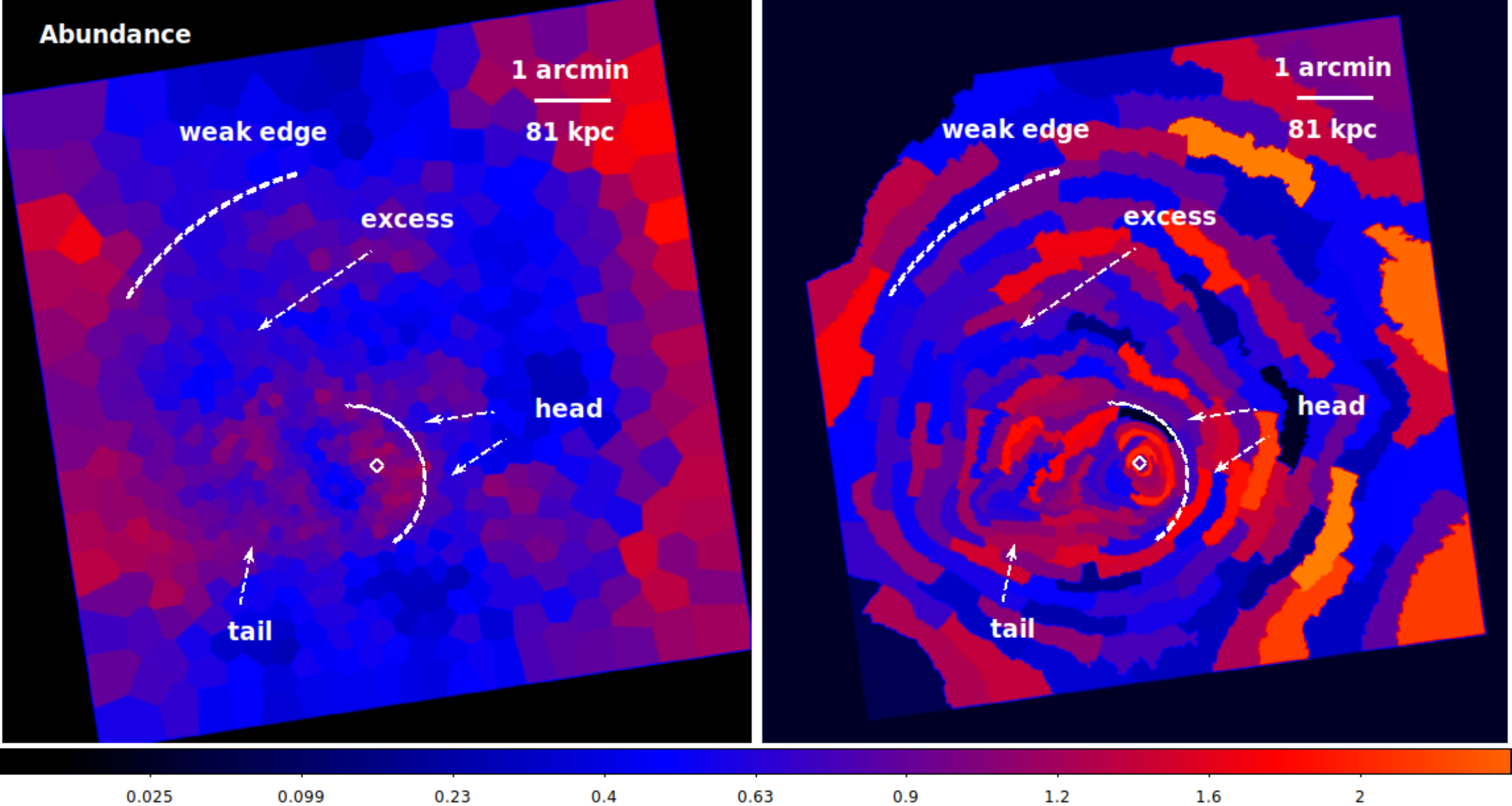}
\caption{Projected \Chandra\ temperature map (top) and abundance map (bottom) generated by the approaches of the improved CVT (left) and the Contour Binning (right), respectively. The position of the X-ray emission peak is marked by `$\lozenge$'. The head, tail, excess, and weak edge are also marked. \label{fig7}}
\end{figure}

The spectra of each bin were fitted with a single APEC model (the absorption fixed at $N_{\rm H} = 3.00 \times 10^{20}$~$\rm cm^{-2}$) and the best-fit model parameters are plotted in Figure~\ref{fig7}. We find that the two approaches yield fairly consistent results. In the X-ray bright core, the radius of which is $\sim 40$~kpc, the gas is relatively cooler ($\sim 3$~keV) and exhibits a slightly higher metallicity ($\sim 1.2~\rm Z_{\sun}$) than in outer regions. This low temperature core region shows an elongated morphology in the north-south direction resembling a mushroom head.
The situation is complicated beyond the central 40~kpc. To the west of the core, there exists an arc where the gas temperatures increase rapidly to $\sim 6$~keV and the metallicity drops to $\sim 0.3-0.8~\rm Z_{\sun}$. The inner edge of the arc closely coincides with the head identified in Figure~\ref{fig1} (see also \S3.2.1).
To the east of the core a strip of cold gas, where the gas temperature ($\sim 3$~keV) is consistent with that of the core but is lower than those of the adjacent regions by $\sim 1-2$~keV (90\% confidence level), extends eastwards to 200~kpc coinciding with the X-ray tail revealed in the X-ray image (Fig.~\ref{fig1}). 
At about $2\arcmin-4\arcmin$ northeast of the core, the gas is cold ($\sim 3$~keV) forming a spiral-like temperature substructure. The position of this substructure is consistent with the region showing the diffuse X-ray emission excess as marked in the residual map (Fig.~\ref{fig1}; see also \S3.3.3). Across the possible weak edge northeast of the excess, the gas temperature shows a mild outward increase by about 1~keV. 
Considering that the X-ray surface brightness also varies mildly across this edge (\S3.2.3), we are precluded from determining whether or not the feature is a cold front.

%------------------------------------------------------------------------------------
% Figure 7
%-----------------------------------------------------------------------------------

\begin{figure}
\epsscale{1}
%\graphicspath{{figure/}}
\centering
\includegraphics[scale=.4]{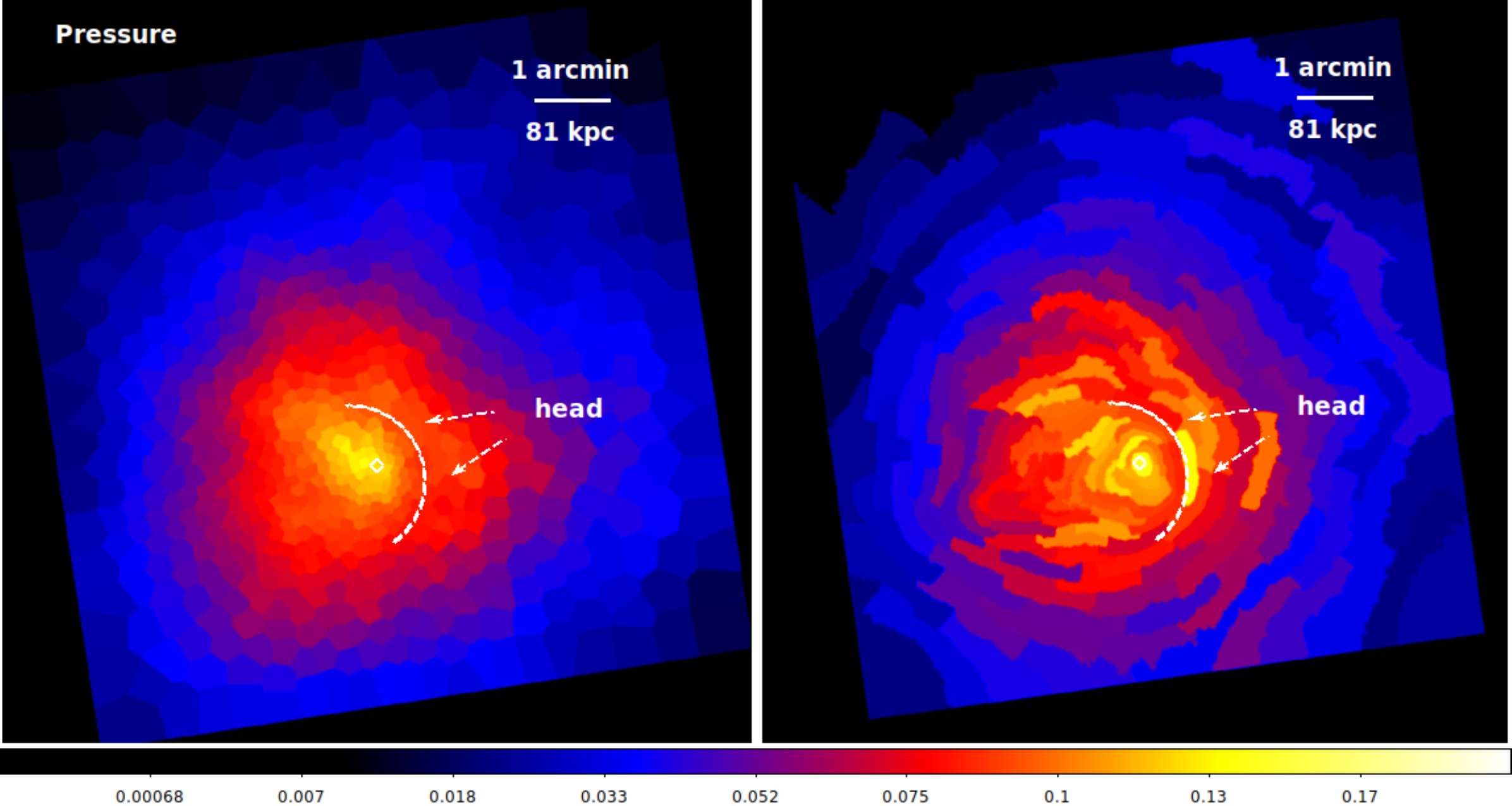}
\includegraphics[scale=.4]{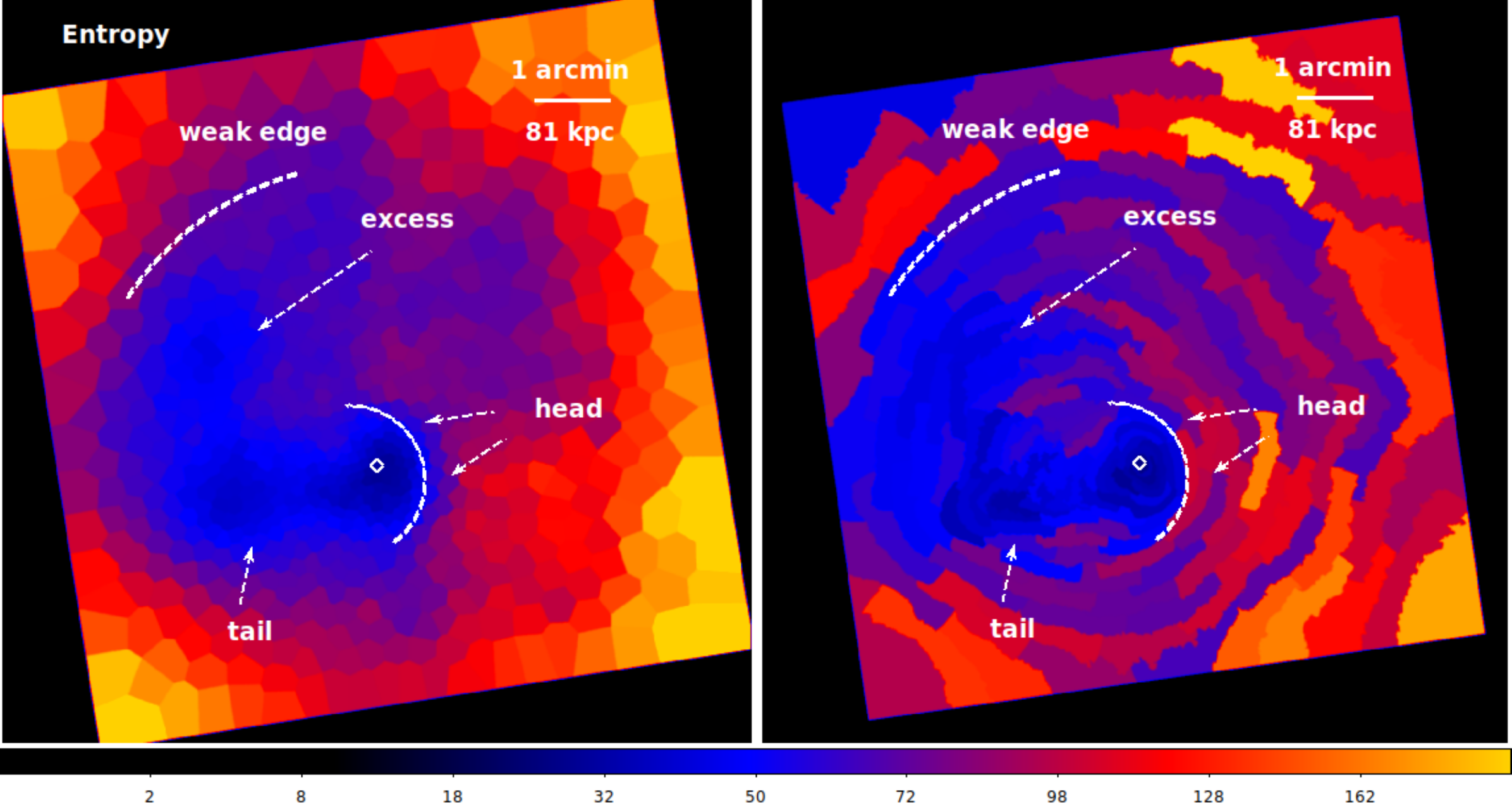}
\caption{Same as Figure.~\ref{fig7} but for projected \Chandra\ pressure map (top) and entropy map (bottom) generated by the methods of the improved CVT (left) and the Contour Binning (right), respectively. \label{fig8}}
\end{figure}

In Figure~\ref{fig8}, we show the corresponding pseudo-pressure and pseudo-entropy maps, which were derived by applying the definitions $P = kT \times \rm EM^{1/2}$ and $S = kT \times \rm EM^{-1/3}$, respectively, where $kT$ is the fitted projected temperature of the temperature map and $\rm EM$ is the projected emission measure (i.e., the APEC normalization divided by the cell's area; \citealt{botteon18,EH19}). 
We find that the pseudo-pressure distribution is relatively symmetric, showing a smooth transfer across the X-ray head and tail, which reflects the fact that the gas dynamics is dominated by gravity. 
This is similar to what was found by \citet{botteon18}, who analyzed 15 galaxy clusters and concluded that the gas pressure across the cold front is almost continuous, as opposed to the prominent pressure jump associated with the shock fronts.
The pseudo-entropy maps, on the other hand, show features similar to those seen in the temperature maps, such as the low-entropy core, tail, and the region that shows the X-ray emission excess, possibly indicating that there is no significant heating from the central AGN activity.  

%-----------------------------------------------------------------------------------
% Figure 8

%------------------------------------------------------------------------------
\subsection{Azimuthally Averaged Spectral Analysis with \XMM\ Data}
%-----------------------------------------------------------------------------

\begin{figure}
\epsscale{1.0}
%\graphicspath{{figure/}}
\centering
\includegraphics[scale=.35, angle=0]{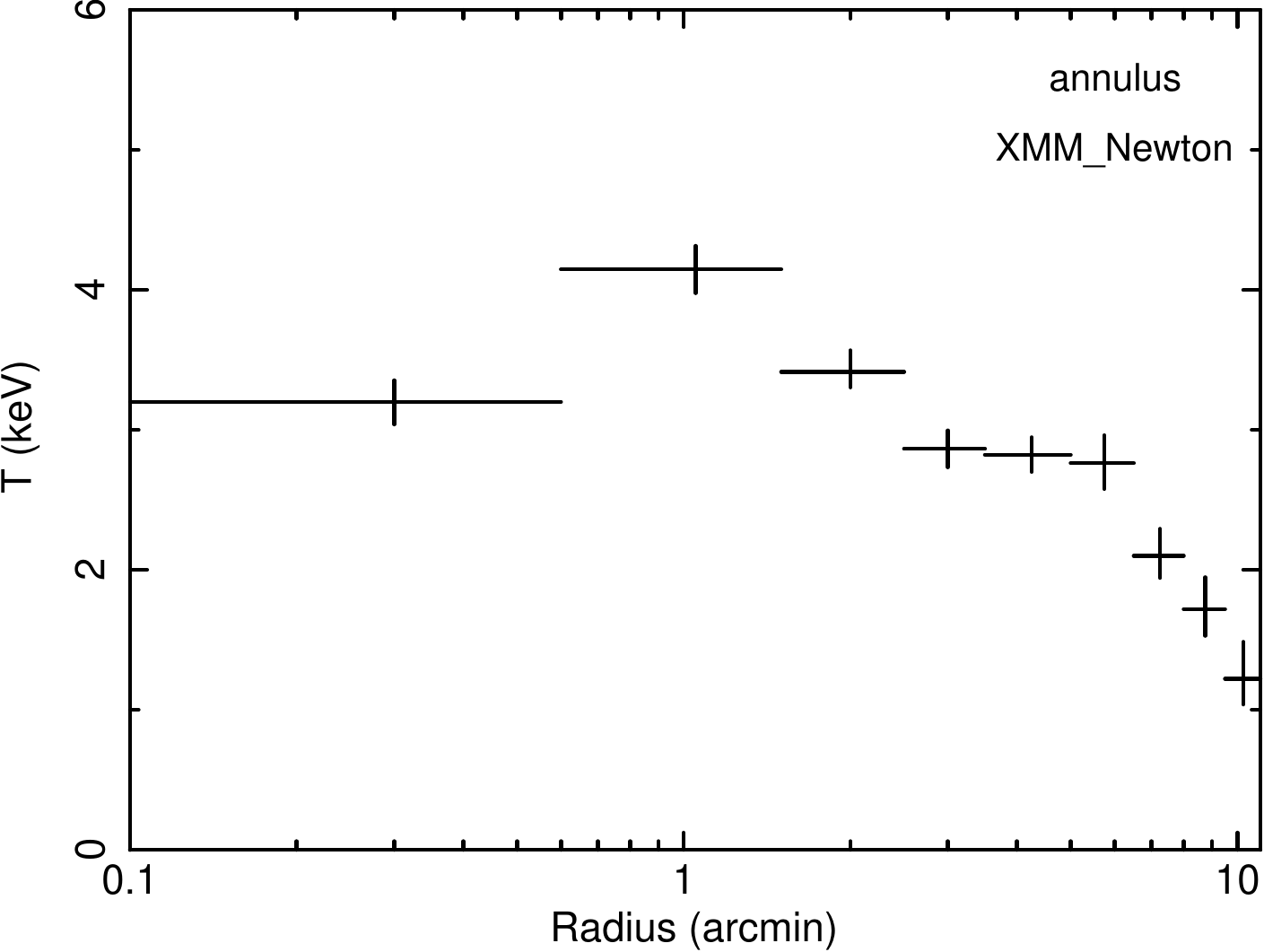}
\includegraphics[scale=.35, angle=0]{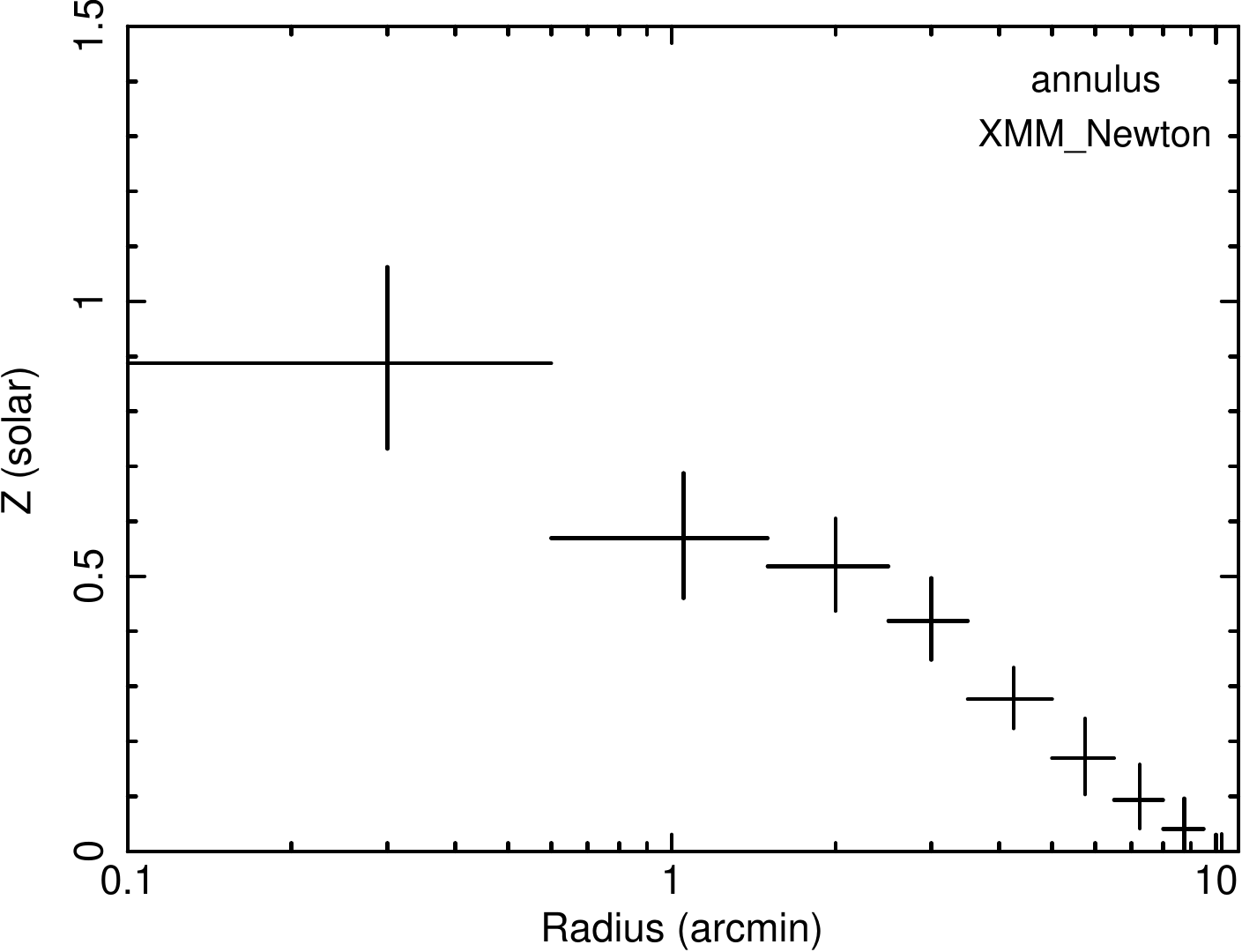}
\includegraphics[scale=.35, angle=0]{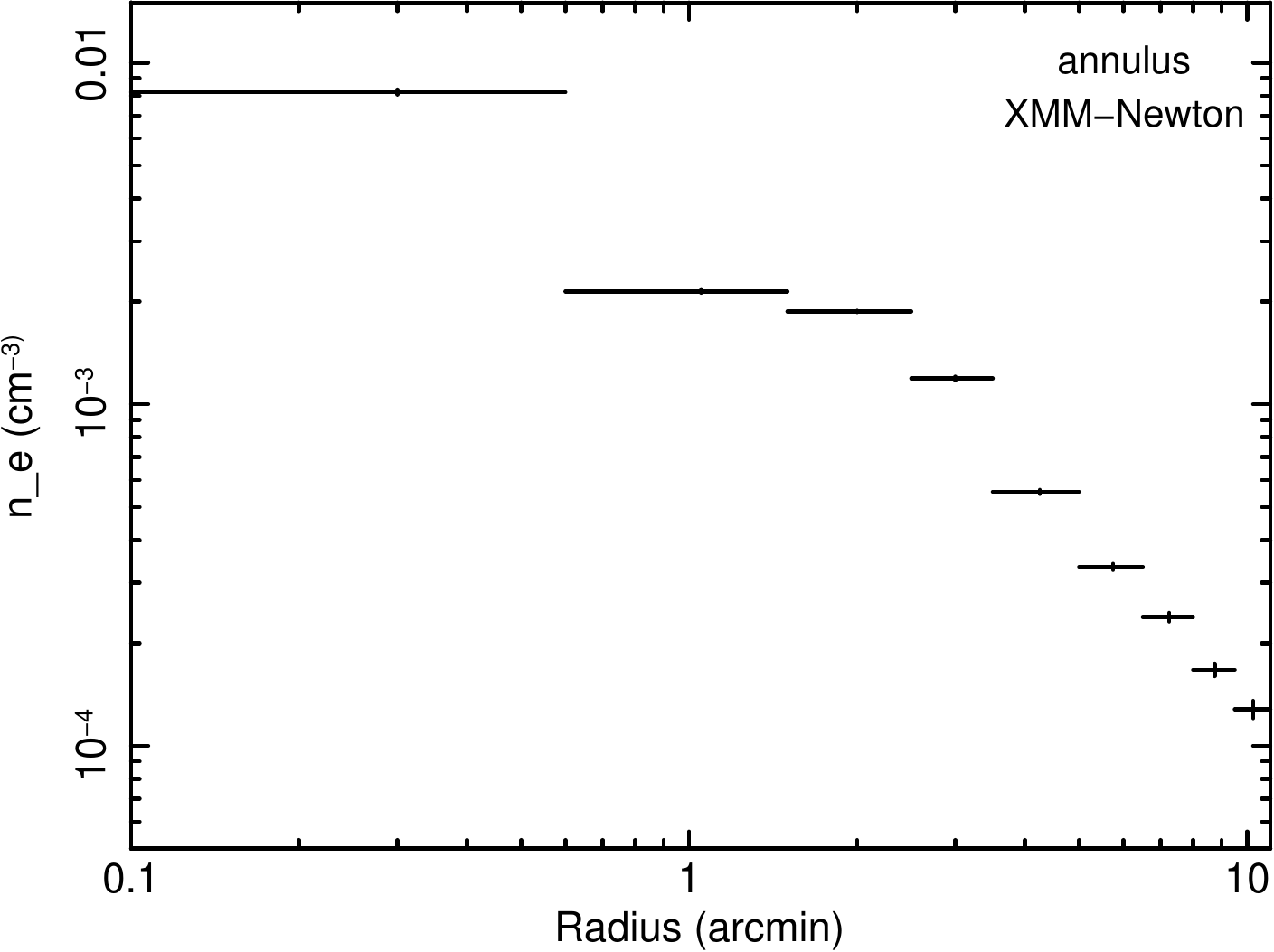}
\caption{Azimuthally averaged projected gas temperature (top) and metal abundance (middle) profiles derived by fitting the \XMM\ EPIC spectra. The corresponding electron number density distribution is shown in the bottom. \label{fig9}}
\end{figure}

In order to calculate the mass distributions in Abell~1775 (\S3.5), we produced the projected gas temperature and metal abundance profiles (Fig.~\ref{fig9}), which extend to the outskirt region (11\arcmin\ or 891~kpc, which is about $1r_{\rm 500}$\footnote{~$r_{\rm 500}$ is defined as the radius within which the mean mass density of the system is 500 times the critical density of the universe at the system's redshift, and $M_{\rm 500}$ is defined as the total gravitating mass within $r_{\rm 500}$.}), by fitting the \XMM\ EPIC spectra extracted from a set of annular regions (the region covered by the \Chandra\ ACIS instrument is not large enough to obtain an overall description of the gas properties). 
We used SAS tasks \texttt{mos-spectra} and \texttt{pn-spectra} to extract the spectra and create the ARFs and RMFs for the EPIC-MOS and PN data, respectively. In order to correct the effect caused by the scattering of photons, which is energy-dependent and is described by the finite point-spread function (PSF) of the detector, we employed the SAS task \texttt{arfgen} to modify the corresponding ARFs \citep{snowden08}. Since the widths of annuli used here are larger than the PSF sizes, the difference between the corrected and uncorrected temperatures is found to be less than 6\% even in the innermost region.
The corresponding electron number density distribution (Fig.~\ref{fig9}) was estimated from the normalization of the APEC model $10^{-14} \int n_{\rm e}n_{\rm H}dV / (4\pi[D_{\rm A}(1+z)]^{2}$), where $D_{\rm A}$ is the angular diameter distance to the source, and $n_{\rm e}$ and $n_{\rm H}$ ($=0.82n_{\rm e}$) are the electron and hydrogen nucleus densities, respectively.
%--------------------------------------------------------------------------
% Figure 9
%-------------------------------------------------------------------------- 
 
We find that the azimuthally averaged projected gas temperature shows a small increase from $3.20_{-0.16}^{+0.15}$~keV in the innermost region ($< 0\arcmin.6$ or 48.6~kpc) to $4.15\pm0.17$~keV at $\sim 1\arcmin$ ($\sim 81$~kpc), and a gradual descent in $\gtrsim 2\arcmin$, which agrees with the temperature profile obtained with the \Chandra\ ACIS data in the W sector (\S3.2). This profile is similar to those obtained in other galaxy clusters, which typically show an outward temperature increase in the inner regions to a peak value at $\sim 0.2r_{\rm 500}$ and then a continuous temperature drop out to about $r_{500}$ \citep{zhu16}, implying that within about $0.2r_{\rm 500}$ the thermodynamic effect of the gravity has been greatly modified by the effects of non-gravitational processes (essentially radiative cooling and AGN feedback), while in outer regions the gravity dominates the gas thermodynamics \citep{zhu21}. 
The azimuthally averaged projected metal abundance shows a significant peak of $0.89_{-0.16}^{+0.18}$~$\rm Z_{\sun}$ in the innermost region ($< 0\arcmin.6$ or 48.6~kpc) and a monotonous outward decrease to a very low value ($\sim 0.1$~$\rm Z_{\sun}$) in the outermost annuli. Similar to the metal abundance, the electron number density also shows a central excess in the $< 0\arcmin.6$ region, where the central gas density reaches $8.20\pm0.14 \times 10^{-3}$~$\rm cm^{-3}$ that is close to the core densities of typical weak cool core (WCC) clusters \citep{hudson10,zhang16}. Using the derived gas temperature and electron number density, we have estimated the gas cooling time of the core region using the following equation \citep{sarazin88}
\begin{equation}
t_{\rm cool} = 8.5 \times 10^{10}~{\rm yr} \left(\frac{n_{\rm H}}{10^{-3}~{\rm cm^{-3}}}\right)^{-1} \left(\frac{T}{10^{8}~ \rm K}\right)^{1/2},
\end{equation}
and obtained a cooling time of about $5.26$~Gyr. Combining these results, the cluster can be classified as a WCC, a category that has not reached the final relaxed state possibly due to disturbances caused by a recent merger or extinct AGN outbursts \citep{hudson10}.

%------------------------------------------------------------------------------
\subsection{Spatial Distributions of Gas Mass and Total Gravitational Mass}
\begin{figure}
\epsscale{1.0}
%\graphicspath{{figure/}}
\centering
\includegraphics[scale=.5]{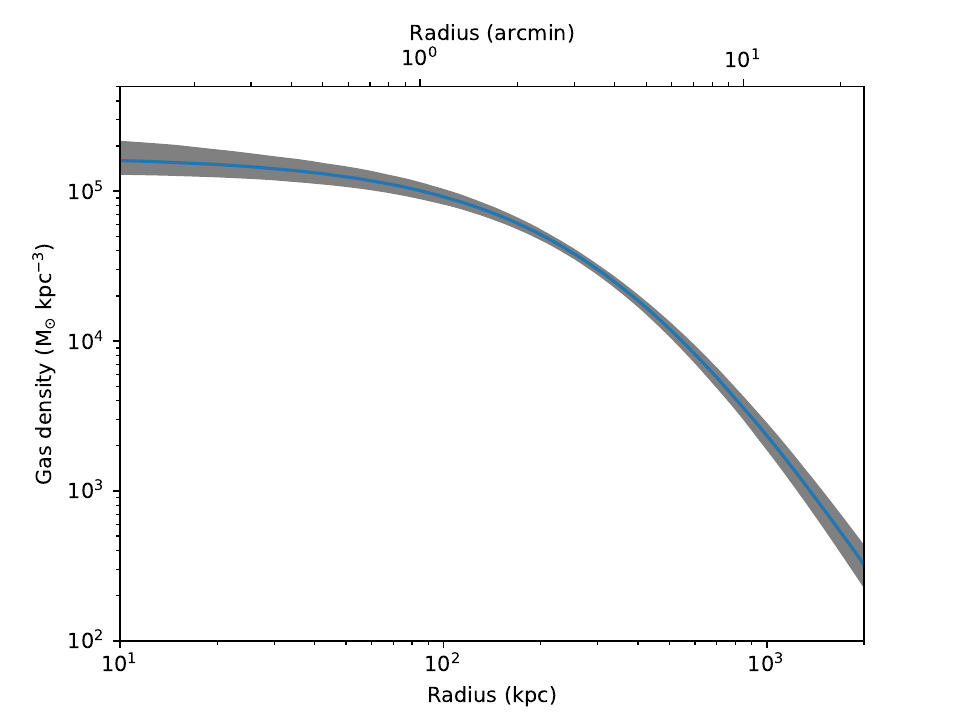}
\includegraphics[scale=.5]{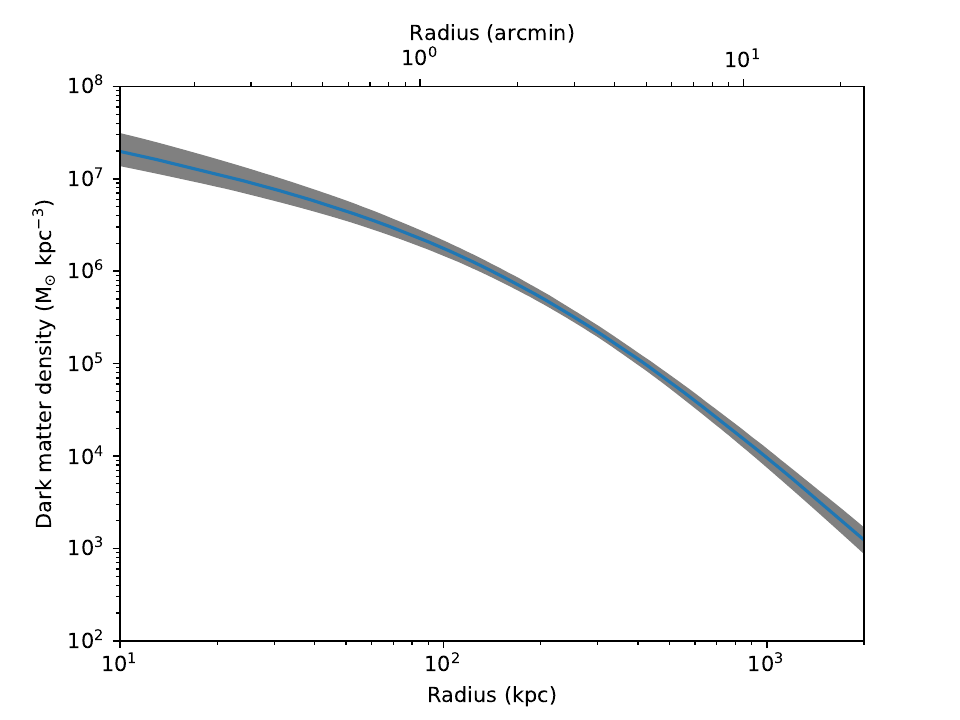}
\includegraphics[scale=.5]{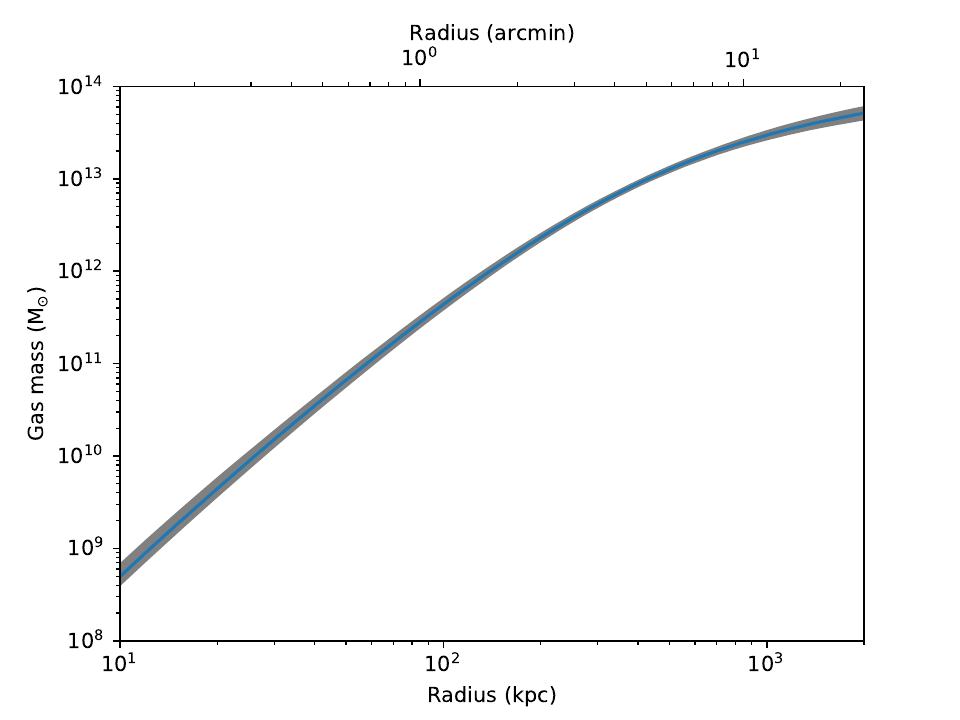}
\includegraphics[scale=.5]{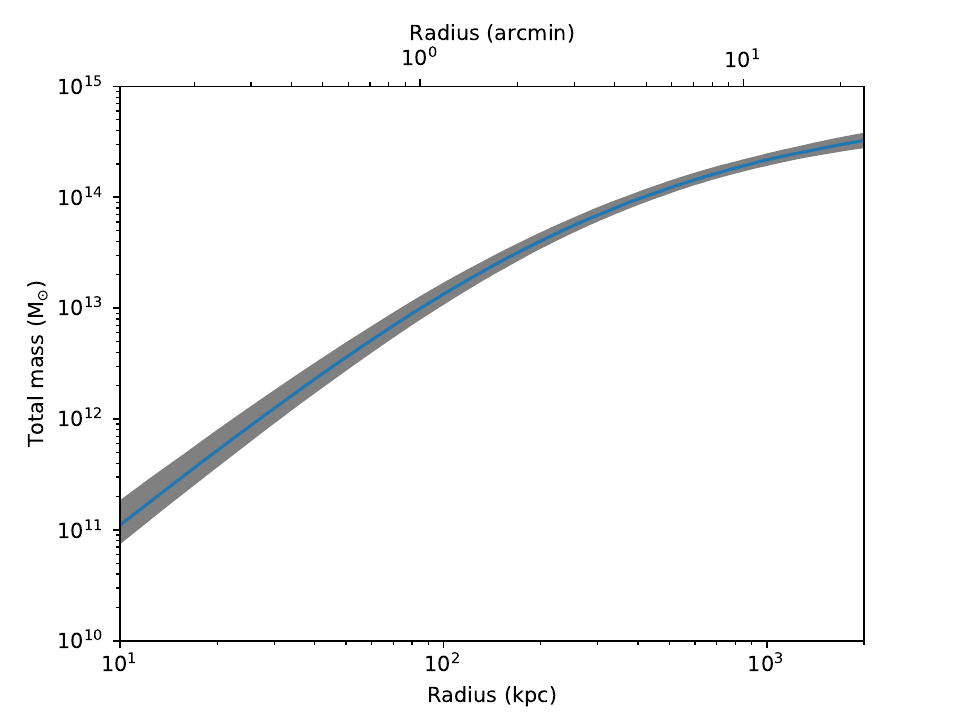}
\caption{Distributions of the gas density (top-left), dark matter density (top-right), gas mass (bottom-left), and total gravitational mass (bottom-right), which are calculated by applying a revised version of the thermodynamical ICM model provided by \citet{zhu16,zhu21}. The shaded areas represent the 90\% uncertainty ranges. \label{fig10}}
\end{figure}

%-----------------------------------------------------------------------------
The gas and total gravitational masses of Abell~1775 can be determined by applying a revised thermodynamical ICM (RTI) model provided by \citet{zhu16,zhu21}, which takes into account the effects of both gravitating and non-gravitational processes (e.g., feedback and radiation). In the RTI model, the total gravitational mass is described by using the generalized NFW profile \citep{NFW97}
\begin{equation}
\rho(r)=\frac{\rho_0}{(r/r_s)^{\delta_1}(1+r/r_s)^{\delta_2-\delta_1}},
\end{equation}
where $r_{\rm s}$ is the scale radius, $\rho_{\rm 0}$ is the critical density, and $\delta_{\rm 1}$ ($\delta_{\rm 2}$) is the inner (outer) slope. 
Meanwhile, the gas density profile $\rho_{\rm gas}(r)$ is described with the model proposed by \citet{PL15}
\begin{equation}
\rho_{\rm gas}(r)=\Gamma f_{\rm gas}\left(\frac{r}{r_{\rm shock}}\right)^{3\Gamma-3}\rho(r_{\rm shock}\left[\frac{r}{r_{\rm shock}}\right]^{\Gamma}),
\end{equation}
where $r_{\rm shock}$ is the radius at which virial shock happens, $\Gamma$ is the gas density jump at the $r_{\rm shock}$, $f_{\rm gas}$ is the average gas fraction within $r_{\rm shock}$, and $\rho(r)$ is the profile of the total gravitational mass density profile mentioned above. After fitting the observed gas temperature and electron number density profiles via Bayesian statistics, which uses the probability to quantify the uncertainty in the model fittings (for details of the Bayesian statistics, see \citealt{AH13} and \citealt{zhu21}), we obtain the best-fit parameters which are listed in Table~\ref{tbl-4}, and calculate the distributions of gas, dark matter, and total gravitational mass with 90\% confidence level (Fig.~\ref{fig10}). Based on these, we obtain $r_{\rm 200}=1266_{-59}^{+67}$~kpc, $r_{\rm 500}=852_{-35}^{+43}$~kpc, $M_{\rm tot, 200}=2.53_{-0.35}^{+0.42} \times 10^{14}$~$\rm M_{\sun}$ and $M_{\rm tot, 500}=1.93_{-0.22}^{+0.30} \times 10^{14}$~$\rm M_{\sun}$, and list the best-fit results in Table~\ref{tbl-5}. The obtained $r_{\rm 200}$ and $M_{\rm 200}$ are consistent with the result of \citet{KK09}, who estimated that the dynamic mass of Abell~1775 within $r_{\rm 200}$ ($\approx 1.39$~Mpc) is $3.28\pm1.25 \times 10^{14}$~$\rm M_{\sun}$ by adopting the measurement of the cluster velocity dispersion in the optical band. These results are adopted to approximate the initial profiles of the main cluster in our numerical studies to be presented in \S4.2.2. 

%--------------------------------------------------------------------------
% Figure 10

%--------------------------------------------------------------------------
% Table 4
\begin{deluxetable}{ccccccc}
\tabletypesize{\footnotesize}
\tablewidth{0.pt}
\centering
\tablecolumns{7}
\tablecaption{Best-fit model parameters for the total gravitational mass and electron number density distributions obtained with the RTI model \citep{zhu21}. \label{tbl-4}}
\tablehead{
\colhead{$r_{\rm s}$}  &  \colhead{$\rho_{\rm 0}$}  &  \colhead{$\delta_{1}$}  &  \colhead{$\delta_{2}$}   &\colhead{$r_{\rm shock}$} &  \colhead{$\Gamma$}  &\colhead{$f_{\rm gas}$}\\
\colhead{(kpc)}  &  \colhead{($\rm M_{\sun}~kpc^{-3}$)}  &  \colhead{}  &  \colhead{}  &  \colhead{(kpc)}  &  \colhead{}   &  \colhead{} }
\startdata
   $224\pm33$  &   $3\pm0.8 \times 10^{6}$  &  $0.66\pm0.13$   &   $3.32\pm0.22$   &  $2486\pm789$  &  $1.26\pm0.05$  &  $0.16\pm0.02$  \\
\enddata
\end{deluxetable}
%--------------------------------------------------------------------------
% Table 5
\begin{deluxetable}{cccccccc}
\tabletypesize{\footnotesize}
\tablewidth{0.pt}
\centering
\tablecolumns{8}
\tablecaption{Best-fit results of Abell~1775. \label{tbl-5}}
\tablehead{
\colhead{$r_{\rm 500}$}  &  \colhead{$r_{\rm 200}$}  &  \colhead{$M_{\rm gas, 500}$}  &  \colhead{$M_{\rm tot, 500}$}   &  \colhead{$M_{\rm gas, 200}$}    &  \colhead{$M_{\rm tot, 200}$} & \colhead{$f_{\rm gas, 500}$} & \colhead{$f_{\rm gas, 200}$} \\
\colhead{(kpc)}  &  \colhead{(kpc)}  &  \colhead{($\rm 10^{14}~M_{\sun}$)}   &  \colhead{($\rm 10^{14}~M_{\sun}$)}  &  \colhead{($\rm 10^{14}~M_{\sun}$)}  &  \colhead{($\rm 10^{14}~M_{\sun}$)} &  \colhead{}  &  \colhead{}  }
\startdata
 $852_{-35}^{+43}$  &  $1266_{-59}^{+67}$   &  $0.25\pm0.03$  &  $1.93_{-0.22}^{+0.30}$   &  $0.37\pm0.05$  &  $2.53_{-0.35}^{+0.42}$  &   $0.13\pm0.01$ &  $0.15\pm0.02$  \\
\enddata
\end{deluxetable}

\clearpage

%===================
\section{DISCUSSION}
%===================
%===========================================
\subsection{Motion of the Gas Core and the Induced Kelvin-Helmholtz Instabilities}
%===========================================
%===========================================
\subsubsection{Velocity of the Cold Front}
%===========================================

When a cold dense gas core moves through a hot ICM, both of which are originally undisturbed, a steady cold front will form due to the ram pressure exerted on the leading edge of the gas core (i.e., the gas pressure difference across the edge). 
Given the ratio of the thermal pressure at the stagnation point ($p_{\rm 0}$) to that in the free stream region ($p_{\rm 1}$), the velocity of gas core can be constrained in terms of Mach number using the following expression \citep{LL59,vikhlinin01}
\begin{equation}
\frac{p_{\rm 0}}{p_{\rm 1}} = 
 \begin{cases}
 	\left( 1 + \frac{\gamma-1}{2} \mathcal{M}^{2} \right) ^ {\frac{\gamma}{\gamma - 1}}  & \quad \rm{if}~\mathcal{M} \leqslant 1 \\
 	\left(\frac{\gamma+1}{2} \right)^ {\frac{\gamma + 1}{\gamma - 1}}  \mathcal{M}^{2} \left(\gamma - \frac{\gamma-1}{2 \mathcal{M}^{2}} \right) ^ {\frac{-1}{\gamma - 1}}  & \quad \rm{if}~\mathcal{M} > 1
 \end{cases},
\end{equation}
where $\mathcal{M}=v/c_{s}$ is the Mach number of the cold front ($v$ is the gas core's velocity and $c_{s}$ is the sound speed), ${\gamma}=5/3$ is the adiabatic index assuming an ideal gas.
Since the gas pressure in the stagnation region ($p_{\rm 0}$) is equal to that inside the front \citep{vikhlinin01,MV07}, we calculate it using the gas temperature $T_{\rm 0}$ and density $n_{\rm e,0}$ measured in the innermost pie region ($r<0\arcmin.6$) in the W sector. In the cases when a shock precedes the cold front, the determination of the gas pressure in the free stream is somewhat delicate because the selection of a free stream becomes difficult due to the turbulent wake of the shock. In this work, however, since no evidence is found for the existence of a shock front beyond $\sim 0\arcmin.6$ in the W sector (\S3.2.1), we simply take the second pie region in the W sector ($0\arcmin.6-1\arcmin.2$) as the free stream region and use the gas temperature $T_{\rm 1}$ and density $n_{\rm e,1}$ measured therein to estimate $p_{\rm 1}$; the same choice was made by, e.g., \citet{osullivan19} who argued that it is difficult to determine a proper region that can describe the free stream accurately and the results for different choices are comparable. Thus, we obtain $p_{\rm 0}/p_{\rm 1} = 1.61_{-0.16}^{+0.21}$ and $\mathcal{M} = 0.79 \pm 0.02$, which implies a subsonic velocity of the core $v=\mathcal{M}c_{s}=\mathcal{M}(\gamma T_{\rm out}/m_{\rm p}\mu)^{1/2} = 937_{-44}^{+53}$~$\rm km~s^{-1}$, where $m_{\rm p}$ is the proton mass and $\mu = 0.6$ is the mean molecular weight of the ICM. This result agrees with the negative detection of the shock in \S3.2.1. Similar core velocity and Mach number of the cold front have been reported in many galaxy clusters and groups, such as Ophiuchus cluster \citep{million10} and NGC~5338 group \citep{osullivan19,wang19}, and both of two systems presents an arc-shaped edge and a stripped gas tail simultaneously.
Using this velocity we estimate that the time needed to form the gas tail is $\sim 1.7 \times 10^{8}$~yr, a value consistent with the timescale of gas stripping processes in the typical mergers (\citealt{AM06}; see also \S4.2 for the corresponding result derived in our simulations).

%===========================================
\subsubsection{Kelvin-Helmholtz Instabilities}
%===========================================

As shown in Figure~\ref{fig2} the X-ray cold front and the ram pressure stripped gas tail exhibit some signatures that are typical for KHI distortions, including two noses, two wings (or rolls), one striking concave bay, and a split gas tail. Since the growth of such instabilities or turbulence can be suppressed either by the magnetic field parallel to the surface of the cold front, or by the local ICM viscosity, or, sometimes, by both, as shown in many numerical simulations \citep[e.g.,][]{ZML11,roediger12a,roediger13}, we are enabled to use these features as a probe to investigate the physical properties of the gas. For example, by ignoring gas viscosity, we may estimate the upper limit of the magnetic field by applying the stability condition \citep{VM02}
\begin{equation}
\frac{B_{\rm c}^{2}+B_{\rm h}^{2}}{8\pi} > \frac{1}{2} \frac{\gamma \mathcal{M}^{2}}{1+T_{\rm c}/T_{\rm h}} P_{\rm ICM},
\end{equation} 
where $B_{\rm c}$ ($B_{\rm h}$) and $T_{\rm c}$ ($T_{\rm h}$) are the magnetic field strength and gas temperature in the inner (outer), i.e., cold (hot), side of the cold front, respectively, $\mathcal{M}$ is the Mach number of the shear flow, and $P_{\rm ICM}$ is the ambient ICM pressure ($P_{\rm ICM} \sim 0.016$~$\rm keV~cm^{-3}$ in our case; \S3.2.1). When the gas is assumed to be adiabatic ($\gamma = 5/3$), we find that in Abell~1775 the maximum total magnetic field strength at the front is ($B_{\rm c}+B_{\rm h}) \sim$ 11.2~$\mu$G, which is consistent with that measured in Abell~3667 ($\rm 6~\mu G - 14~\mu G$; \citealt{VM02}), a cluster possessing a sharp cold front for which the evidence for the existence of the magnetic field layer at the cold front was first reported.

On the other hand, for a shear layer with a given shear velocity, viscosity can suppress the growth of KHIs on scales below a certain length. Assuming the Spitzer-like viscosity, the critical wavelength above which the KHI can grow, can be calculated following \citet{roediger13}
\begin{equation}
\lambda_{\rm crit} = 30~{\rm kpc} \frac{\rm Re_{\rm crit}}{30} f_{\mu} \left(\frac{U}{400~{\rm km~s^{-1}}}\right)^{-1} \left(\frac{n_{\rm e}}{10^{-3}~{\rm cm^{-3}}}\right)^{-1} \left(\frac{T}{2.4~{\rm keV}}\right)^{5/2},
\end{equation} 
where $\rm Re_{crit}$ is the critical Reynolds number, $f_{\mu}$ is the viscosity suppression factor, $U$ is the shear velocity, and $n_{\rm e}$ and $T$ are the gas electron density and gas temperature of the ambient ICM, respectively. Also following \citet{roediger13}, we adopt a conservative Reynolds number $\rm Re_{crit}~(=10\sqrt{\frac{(n_{\rm e,c}+n_{\rm e,h})^2}{n_{\rm e,c}n_{\rm e,h}}}$) and a standard shear velocity $U$ that corresponds to a Mach number of $\mathcal{M} =$ 0.5 ($U = \mathcal{M}c_{s} \approx 593$~$\rm km~s^{-1}$ for Abell~1775). Since recent hydrodynamic simulations performed by \citet{roediger15}, which were focused on gas-stripping galaxies hosting either inviscid atmospheres or viscosity ($f_{\mu}$ is in the range of $0.001 - 0.1$) atmospheres, indicated that the KHI substructures shown in the inviscid and low viscosity ($f_{\mu}=0.001$) cases are more ragged than those observed in Abell~1775, while when viscosity is high enough ($f_{\mu}=0.1$) the surface of the simulated cold fronts become much smoother than what we have observed in Abell~1775, we tentatively choose an intermediate viscosity of $f_{\mu}=0.01$. Therefore, we find that in Abell~1775 the growth of KHIs can be suppressed by Spitzer-like viscosity on length-scales below 7~kpc, which is consistent with the smallest KHI scale, i.e., the length of the nose ($\simeq 10$~kpc), found in Abell~1775.

%===========================================
\subsection{Merger Scenarios }
%===========================================
%===========================================
\subsubsection{Two Possible Merger Scenarios}
%===========================================

\begin{figure}
\epsscale{1}
%\graphicspath{{figure/}}
\centering
\includegraphics[scale=.5]{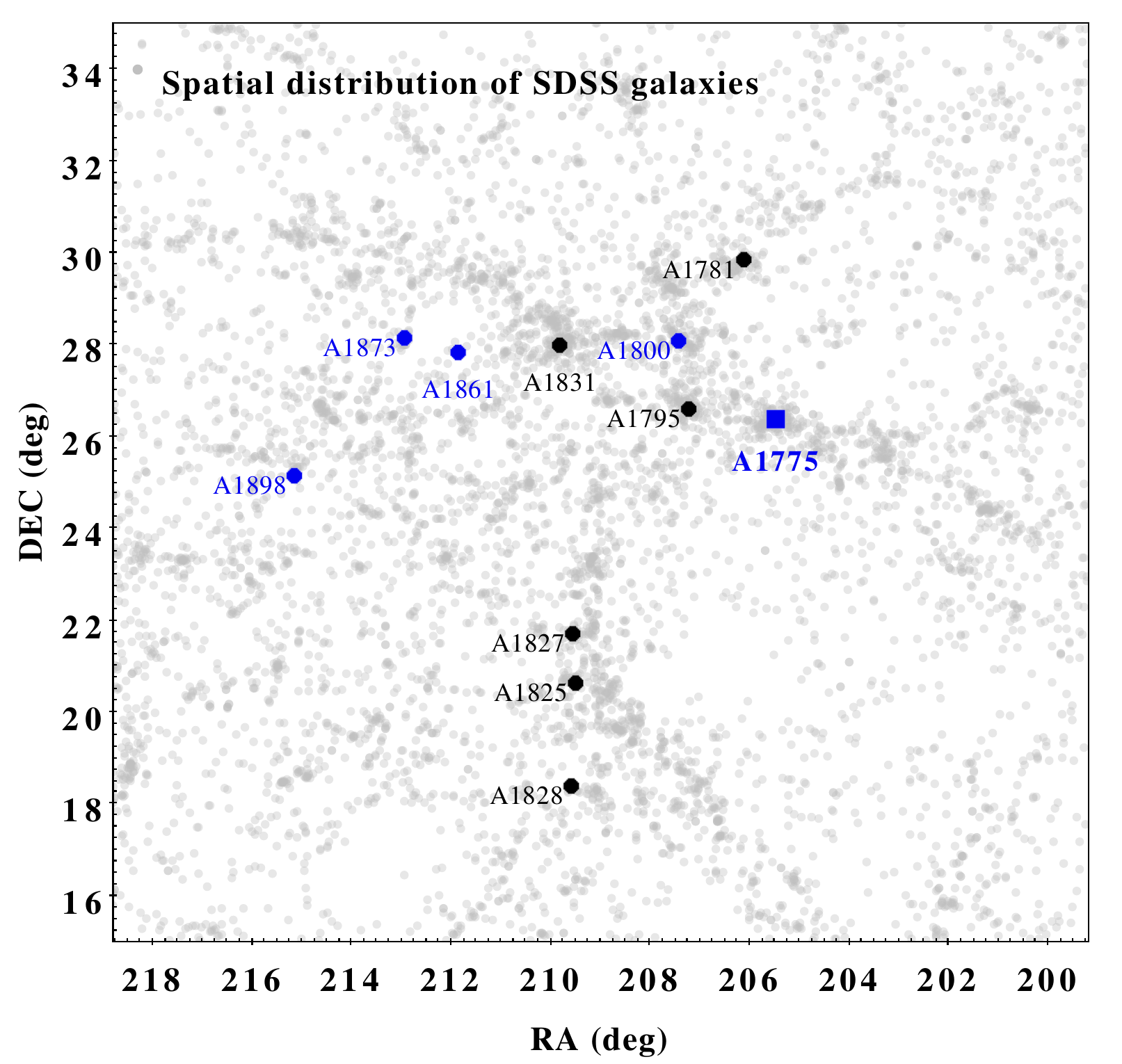}
\caption{The spatial distribution of SDSS galaxies (grey dots) in a redshift range of $0.06<z<0.08$. All the SDSS galaxies are brighter than $r\sim17.7$ mag and have spectroscopic redshift measurements. Major galaxy clusters located in the nearer one (with the mean redshift of 0.061) are marked by black points. Blue markers represent the major galaxy clusters in another supercluster (with the mean redshift of 0.074), and the blue box represents the location of Abell~1775.  \label{fig11}}
\end{figure}

The apparent westward motion of the cluster core, as revealed by its distinctive X-ray morphology, suggests two possible merger scenarios. In the first scenario, Abell~1775 acts as an infalling subcluster and is traveling through a low-density environment toward a more massive system (e.g., another more massive cluster or a supercluster) due to the powerful gravitational attraction of the latter. 
This scenario, however, is not likely to be true, because the nearest massive cluster Abell~1795 is located at a projected distance of $\sim 7.8$~Mpc approximately east of the Abell~1775, and only possesses a total mass of $4.4 \times 10^{14}$~$\rm M_{\sun}$ and a virial radius of about 2~Mpc \citep{bautz09}, which is far from being powerful enough to drag Abell~1775 in the correct direction and form the observed X-ray features. 
Furthermore, by inspecting the SDSS galaxy map (Fig.~\ref{fig11}), we also exclude the possibility that the cold front is formed because Abell~1775 is moving toward the center of the farther one of the Bo\"{o}tes superclusters\footnote{~There are two Bo\"{o}tes superclusters, the mean redshifts of their member clusters are 0.061 and 0.074, respectively. Abell~1775 resides in the farther supercluster.} on a galaxy filament, since the direction of the motion of Abell~1775 is roughly perpendicular to the extension of the filament where Abell~1775 resides in.

%--------------------------------------------------------------------------

% Figure 11
%--------------------------------------------------------------------------

In the second scenario, Abell~1775 is undergoing a two-body merger as the main cluster and is evolving at a stage after the first pericentric passage when the infalling less massive subcluster is moving away, a case that has been intensively studied in the simulations of, e.g., \citet{TH05}, \citet{AM06}, \citet{poole06}, \citet{zuhone11}, and \citet{hu19}. As revealed in these simulations, in major mergers (e.g., mass ratios of 1:1 and 3:1), the cores of both the main cluster and the subcluster can be affected significantly by the ram pressure stripping after the first pericentric passage, resulting in a leading edge and a stripped gas tail. In minor mergers (e.g., 10:1), however, only the core of the subcluster is apparently affected, and usually the ram pressure stripped gas tail cannot develop behind the main core. 
Note that this two-body merger scenario is consistent with the optical evidence, i.e., the existence of two optical subclusters showing a spatial separation of $\sim 1.1$~Mpc and a radial velocity difference of $\sim 2980$~$\rm km~s^{-1}$ \citep{zhang11}. The spatial distribution of the member galaxies also shows a tight correlation with the X-ray emission. In addition to the fact that the peak of the optical main cluster (the BCG UGC~08669 NED01) is within $3\arcsec$ of the X-ray peak (\S3.1.2), it is also found that around the galaxy density peak of the optical subcluster, there exists a very faint X-ray clump, which can be observed in the \ROSAT\ PSPC image. 
We estimate that in $0.1-2.4$~keV the emission of the X-ray subcluster is at least fifteen times fainter than that of the main cluster, corresponding to a luminosity of $L_{\rm X, sub} (\rm 0.1-2.4~keV) \lesssim 2 \times 10^{43}$~$\rm ergs~s^{-1}$. This X-ray subcluster, however, cannot be detected with the \XMM\ EPIC camera, possibly due to the high background level of the EPIC detector as well as the insufficient exposure time.

%===========================================
\subsubsection{Numerical Experiment for the Two-body Merger}
%===========================================

In order to investigate whether or not the second scenario is valid, we have carried out a set of idealized hydrodynamic simulations by using the TreePM-SPH GADGET-2 code \citep{springel05} in a similar manner to that described in \citet{hu19}. In the simulation, the initial profiles of dark matter and gas of the main cluster are set to be the same as the observed ones (\S3.5). The corresponding profiles of the infalling subcluster are scaled down according to the mass ratio between two merging clusters. The mass resolution of the dark matter and gas particles are $\rm 3.6\times 10^{8}~M_{\sun}$ and $\rm 4.7\times 10^{7}~M_{\sun}$, respectively, and the gravitational softening length is 5~kpc. The initial conditions at $t=0$ are as follows: (1) mass ratio $R =M_{\rm 200,mian} / M_{\rm 200,sub} =$ 3, 5, and 10; (2) initial separation $d_{\rm 0} =2\times(r_{\rm 200,main}+r_{\rm 200,sub})$; (3) initial relative velocity $v_{\rm 0} = 200-1000$~$\rm km~s^{-1}$; (4) impact parameter $P=0-1000$~kpc. In addition to these, the gas fraction of the subcluster ($f_{\rm gas,sub}$) is set to within the range of $1\%-10\%$, based on the $L_{\rm X}-M_{\rm total}$ and $M_{\rm total}-f_{\rm gas}$ relations provided by \citet{lovisari15}. We let the merger process occurs in the x-y plane, and the simulation results are projected along the z-axis by using the {\it yt} \footnote{~~https://yt-project.org/} \citep{turk11} to generate the projected X-ray surface brightness and temperature maps.  

By inspecting the simulated maps, we find that in the head-on mergers ($P \simeq 0$), the gas core of the main cluster is always destroyed after the first pericentric passage if the subcluster contains abundant gas ($f_{\rm gas,sub} \geqslant 5\%$). If the subcluster's gas content is low ($f_{\rm gas,sub} \simeq 1\%$), it turns out that the gas substructures in head-on mergers always exhibit symmetrical morphologies, which disagree with the X-ray observations. We find that it is only in the off-axis mergers that the cold gas core of the main cluster has a chance to survive after the subcluster passes by, and its peak may deviate from the dark matter peak by several tens of kpc due to the ram pressure force. As the subcluster moves away from the main cluster, the displaced cold  gas peak of the main cluster starts to turn back and fall toward the bottom of the gravitational well of the main cluster, which is dominated by the dark matter. At this stage, the cold gas core will collide with the outflowing gas that moves in the opposite direction, and the ``mushroom-like'' morphology is likely to be generated in a process similar to the Rayleigh-Taylor instability. Meanwhile, the outermost part of the displaced cold gas, which has experienced obviously adiabatic expansion, may form a plume of gas excess, which always visually connects to the end of the stripped gas tail.

We find that, however, characteristic X-ray substructures (i.e., the sharp leading edge, the gas tail, and spiral emission excess) similar to the observed ones only form when the gas content of the subcluster ($f_{\rm gas,sub} < 5\%$) is low. If the subcluster contains abundant gas ($f_{\rm gas,sub} \geqslant 5\%$), the infalling subcluster can make a significant disturbance in the gas of the main cluster's core during the first pericentric passage, especially when $P$ is small ($< 500$~kpc), making the gas distribution and velocity field more chaotic than the observation. 
This phenomenon is consistent with that result found by \citet{AM06}, who compared the formation of cold fronts with gas-rich and gas-poor subclusters by focusing on the simulations with a mass ratio of 5 and a nonzero impact parameter (e.g., $P=500$~kpc). 

\begin{figure}
\epsscale{1}
%\graphicspath{{figure/}}
\centering
\includegraphics[scale=.28]{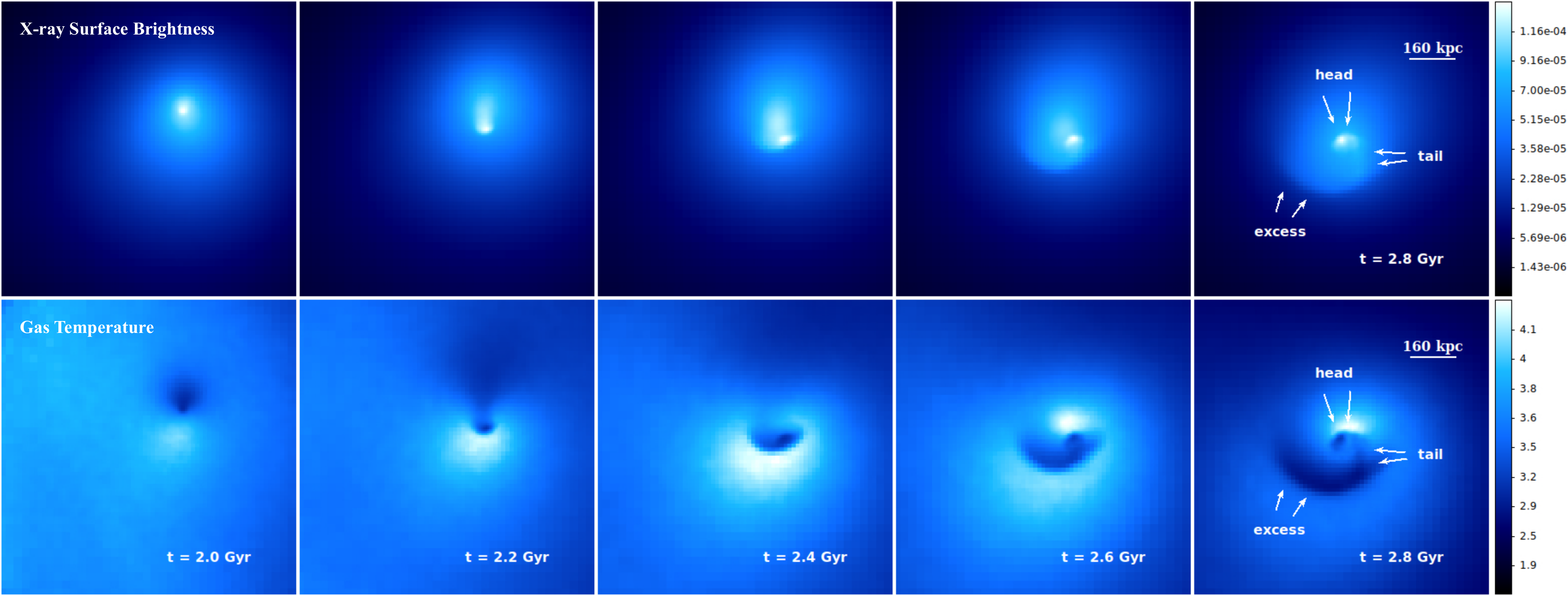}
\caption{The evolution of the X-ray surface brightness ($\rm photons~cm^{-2}~arcmin^{-2}~s^{-1}$) and the gas temperature (keV) in the ``best-match'' model. The characteristic substructures correspond to those identified in the observation are marked on the snapshot at 2.8~Gyr.  \label{fig12}}
\end{figure}

By examining the results of the simulations, we have successfully found a model that best describes the observation, the initial condition configurations of which are $R=5$, $P=500$~kpc, $v_{\rm 0}=500$~$\rm km~s^{-1}$, and $f_{\rm gas,sub}=1\%$. The evolution of the main gas peak predicted with this ``best-match'' model in terms of the X-ray surface brightness and gas temperature maps are presented in Figure~\ref{fig12}. At the snapshot of $t=2.8$~Gyr, the simulated gas properties most resemble the observed ones (Figs.~\ref{fig1} and \ref{fig7}) as described below.
\begin{itemize}
\item Very similar ``mushroom-like'' morphology includes an arc-shaped head and a strip of gas tail. In the simulated X-ray surface brightness map, the radius of the arc-shaped head is about $30-40$~kpc and the length of the tail is about 160~kpc. In addition to these, a spiral-like gas substructure forms an emission excess region located at about 260~kpc to the X-ray peak, and almost vertically connects with the gas tail. 
\item Temperature distribution is similar to the observed pattern. The simulated temperatures of the gas residing in the core and tail are both around $3.5$~keV. In the spiral-like emission excess region the gas temperature is about 2.5~keV.
\item The growth time of the simulated gas tail, which is stripped from the main cluster core, is about 0.2~Gyr (from $t=2.6$~Gyr to $t=2.8$~Gyr). This agrees nicely with the estimation given in \S4.1.1.
\item The best-match mass ratio between the merging subclusters is 5, which agrees well with the results of \citet{KK09}, who estimated the dynamic masses of two subclusters based on the study of the velocity dispersion of member galaxies.
\item In the best-match solution, we find that at about $t=2.8$~Gyr the subscluster has moved outward to over 1~Mpc away from the main cluster's center, roughly in the direction of the tail's extension. This is exactly where the optical subcluster, as well as the faint X-ray halo revealed with \ROSAT, are found \citep{OHF95,KK09,zhang11,MKU89}.
\item \textbf{Assuming isotropic turbulence, we calculated the turbulent velocity \citep{zhuravleva13,beattie19} using the best-match solution, and obtained about 570~$\rm km~s^{-1}$ for the central core region ($< 40$~kpc) and about 1200~$\rm km~s^{-1}$ for the spiral pattern. These results are fairly consistent with that estimated from the radio observations \citep{botteon21}. Combining with the estimated magnetic field of $\sim$ 11.2~$\mu$G at the cold front, which is higher than the typical level expected in the galaxy clusters that possess radio halos, our merger scenario would likely result in a radio halo in the central region.}
\end{itemize}

On the other hand, because the NAT and WAT radio galaxies may possess roughly the same mass as inferred by their similar absolute magnitudes ($-24.51$~mag and $-24.04$~mag, respectively; data from NED), it is natural to speculate that the NAT radio galaxy, whose radial velocity relative to the main cluster is estimated to be $\simeq 1739$~$\rm km~s^{-1}$ \citep{zhang11}, is the central dominating galaxy of the infalling subcluster. However, this is not likely to be true because the simulations of, e.g., \citet{AM06}, \citet{poole06}, \citet{zuhone11}, and \citet{hu19}, show that the core-core interaction in an equal-mass merger is usually violent (even if the collision is off-axis) and will cause some characteristic substructures, such as prominent edges preceding each of the cluster cores or two severely disrupted gas cores, that are absent in Abell~1775. Also, simulations show that when two merging cluster-dominating galaxies approach each other within such a close distance (a projected distance of about 28~kpc), they should be moving approximately in the opposite direction nearly in all cases; in Abell~1775, however, the WAT is moving toward the west and the NAT is heading toward the south. Finally, the NAT is unlikely to associate with the optical subcluster. Therefore, the NAT radio galaxy is more likely to be a single newcomer that is falling into Abell~1775 coincidentally when a two-body merger event occurs.

% \textbf{\red We noticed that a recent discovery of radio halo in the central region of Abell~1775 \citep{botteon21} supports this scenario???.}

%--------------------------------------------------------------------------

% Figure 12

%--------------------------------------------------------------------------

%==================================================
\subsection{Dynamics of the Tailed Radio Galaxies}
%==================================================
Following the method of \citeauthor{douglass08} (2008; see also \citealt{sakelliou08} and \citealt{douglass11}), the shape of the tails (or jets) in tailed radio galaxies, which is bent by ram pressure, can be used to constrain the velocity of the host radio galaxies relative to the surrounding ICM by applying the Euler's equation
\begin{equation}
\frac{\rho_{\rm t}v_{\rm t}^{2}}{r_{\rm c}} = \frac{\rho_{\rm ICM}v_{\rm gal}^{2}}{r_{\rm t}}
\end{equation}
where $\rho_{\rm t}$ is the plasma density in the tail, $v_{\rm t}$ is the velocity of the plasma's bulk motion in the tail, $\rho_{\rm ICM}$ is the ICM density, $v_{\rm gal}$ is the velocity of the host galaxy relative to the ICM, $r_{\rm c}$ is the radius of curvature of the bent tail, and $r_{\rm t}$ is the  cylindrical radius of the tail. 
For the NAT radio galaxy, we adopt $r_{\rm c, NAT}=1.1$~kpc and $r_{\rm t, NAT} = 0.25$~kpc based on the high-resolution 5~GHz map provided in \citet{terni17}; the former of which is consistent with the projected radius of curvature ($\sim 0.8$~kpc) given in \citet{OO85}. 
For the WAT radio galaxy, we obtained $r_{\rm c, WAT}=10$~kpc and $r_{\rm t, WAT} = 0.1$~kpc from the 1.4~GHz map provided in \citet{OO85}. Because both of the two radio galaxies reside in the cluster core ($< 40$~kpc), we approximate the ICM density as $\rho_{\rm ICM} \simeq 8.20\times 10^{-3}$~$\rm cm^{-3}$ as measured within $0\arcmin.6$ (i.e., 48~kpc; \S3.4). 
We apply the jet velocity of 0.2$c$ for the NAT, the value derived by \citet{terni17}, who estimated it based on the analysis of the brightness ratio between the jet and counter-jet. However, since there is no sign of beaming in the WAT jet, we assume that the jet has been decelerated and take a reasonable jet velocity of 0.1$c$ for the WAT \citep{LB02,sakelliou08}. 
We find that, if the tails of the two radio galaxies are assumed to be heavy tails ($\rho_{\rm t}/\rho_{\rm ICM} = 1$; \citealt{guo15}), the derived galaxy velocities are $2.8 \times 10^{4}$~$\rm km~s^{-1}$ and $3 \times 10^{3}$~$\rm km~s^{-1}$ for the NAT and WAT, respectively. If the tails are assumed to be light ($\rho_{\rm t}/\rho_{\rm ICM} = 0.01$; \citealt{guo15}), the derived galaxy velocities are 2800~$\rm km~s^{-1}$ and $300$~$\rm km~s^{-1}$ for the NAT and WAT, respectively.
Apparently, the latter is favored by the results of \citet{bliton98}, who analyzed a sample of NAT radio galaxies and pointed out that the relative velocity of their host galaxies are always $1-3$ times the cluster velocity dispersion (generally within hundreds to several thousands of $\rm km~s^{-1}$), and is also consistent with the general view of the WAT, which sites roughly at the center of the gravitational well's bottom with a low peculiar velocity ($\sim 300$~$\rm km~s^{-1}$; \citealt{burns81}).

%==============================
\section{CONCLUSIONS}
%==============================

By analyzing the high-quality \Chandra\ and \XMM\ archive data and running a series of idealized hydrodynamic simulations, we obtain the following results:
\begin{enumerate}
	\item The X-ray emission of Abell~1775 exhibits an arc-shaped edge (i.e., head) at $\sim48$~kpc west of the X-ray peak, a split cold gas tail that extends eastward to $\sim 163$~kpc, and a plume of spiral-like X-ray excess (within about $81-324$~kpc northeast of the cluster core) that connects with the end of the tail. The head, across which the projected gas temperature rises from $3.39_{-0.18}^{+0.28}$~keV to $5.30_{-0.43}^{+0.54}$~keV, appears to be consistent with a cold front with a Mach number $\mathcal{M} = 0.79 \pm 0.02$.
	\item  Along the surfaces of the cold front and tail, several KHI distortions, including two noses, two wings (or rolls), one striking concave bay, and a split tail, are observed and used to constrain the physical properties of the ICM. The upper limit of the magnetic field is estimated to be about 11.2~$\mu$G and the viscosity suppression factor is found to be about 0.01.
	\item Based on the X-ray results and the bimodal radial velocity distribution of the member galaxies in the optical band, a two-body merger scenario is proposed. We carry out a set of idealized hydrodynamic simulations by using the GADGET-2 code and find that the observed X-ray emission and temperature distributions can be best reproduced in a merger with a mass ratio of 5 after the first pericentric passage at the snapshot of $t=2.8$~Gyr. \textbf{Moreover, our merger scenario with turbulent velocity of $\sim 570$~$\rm km~s^{-1}$ in the central region would likely result in a radio halo in the central region \citep{botteon21}.} 
	\item The NAT radio galaxy is more likely to be a single galaxy falling into the cluster center at the relative velocity of 2800~$\rm km~s^{-1}$, as constrained by its radio morphology and the light jet model.
\end{enumerate}

%============================
\acknowledgments
%============================
\textbf{We sincerely thank the referee for providing valuable comments.} This work is supported by the Ministry of Science and Technology of China (grant No. 2018YFA0404601), and the National Science Foundation of China (grant Nos. 11621303, 11835009 and 11973033).

\clearpage

%===============================================================================================

\clearpage

%-----------------------------------------------------------------------------------------------

\end{document}